\documentclass[sigconf]{acmart}

\usepackage{graphicx}
\usepackage{xspace}
\usepackage{xargs}
\usepackage{balance} 
\usepackage{enumerate}
\usepackage{amsmath}
\usepackage{xcolor}
\usepackage{ifthen}
\usepackage{soul}
\usepackage{caption}
\usepackage{subcaption}
\usepackage{url}
\usepackage[normalem]{ulem}
\usepackage{booktabs}
\usepackage{multirow}
\usepackage{color, colortbl}

\usepackage[labelformat=simple]{subcaption}

\newcommand{\blue}[1]{\textcolor{black}{#1}}

\DeclareMathOperator*{\argmin}{argmin}

\newcommand{\code}[1]{\texttt{#1}}
\newcommand{\topic}[1]{\smallskip
\noindent\textbf{#1.}}

\newcommand{\subtopic}[1]{\smallskip
\noindent\underline{\em #1.}}

\newcommand{\squishlist}{
   \begin{list}{$\bullet$}
    { \setlength{\itemsep}{0pt}
      \setlength{\parsep}{2pt}
      \setlength{\topsep}{6pt}
      \setlength{\partopsep}{0pt}
      \leftmargin=10pt
\rightmargin=0pt
\labelsep=5pt
\labelwidth=10pt
\itemindent=0pt
\listparindent=0pt
\itemsep=\parsep
    }
}
\newcommand{\squishend}{\end{list}} 

\newcommand{\fig}{figures}

\setcopyright{rightsretained}
\renewcommand\footnotetextcopyrightpermission[1]{}
\makeatletter
\def\@copyrightspace{\relax}
\def\@copyrightblurb{\relax}
\def\@mkbibcitation{\relax}
\makeatother

\begin{document}

\copyrightyear{2021}
\acmYear{2021}
\acmConference[SIGMOD '21]{Proceedings of the 2021 International Conference on Management of Data}{June 20--25, 2021}{Virtual Event, China}
\acmBooktitle{Proceedings of the 2021 International Conference on Management of Data (SIGMOD '21), June 20--25, 2021, Virtual Event, China}\acmDOI{10.1145/3448016.3457566}
\acmISBN{978-1-4503-8343-1/21/06}


\title[Production Machine Learning Pipelines: Empirical Analysis and Optimization Opportunities]{
Production Machine Learning Pipelines:\\
Empirical Analysis and Optimization Opportunities}

\settopmatter{authorsperrow=4}

\author{Doris Xin}
\authornote{Work done while at Google.}
\affiliation{%
  \institution{University of California, Berkeley}
  \city{Berkeley}
  \state{California}
  \country{USA}
}
\email{dorx@berkeley.edu}

\author{Hui Miao}
\affiliation{%
  \institution{Google}
  \city{Mountain View}
  \state{California}
  \country{USA}
}
\email{huimiao@google.com}

\author{Aditya Parameswaran}
\affiliation{%
  \institution{University of California, Berkeley}
  \city{Berkeley}
  \state{California}
  \country{USA}
}
\email{adityagp@berkeley.edu}

\author{Neoklis Polyzotis}
\affiliation{%
  \institution{Google}
  \city{Mountain View}
  \state{California}
  \country{USA}
}
\email{npolyzotis@google.com}

\begin{abstract}

Machine learning (ML) is now commonplace,
powering data-driven applications in 
various organizations.
Unlike the traditional perception of ML
in research, 
ML production pipelines are complex,
with many interlocking analytical 
components beyond training,
whose sub-parts are 
often run multiple times on overlapping subsets of data.
However, there is a lack of quantitative evidence
regarding the lifespan, architecture, frequency,
and complexity
of these pipelines to understand how
data management research can be used
to make them more efficient, effective, robust,
and reproducible.
To that end, we analyze the provenance graphs of 3000
production ML pipelines at Google, comprising over 450,000 models trained, 
spanning a period of over four months,
in an effort to understand the complexity and challenges
underlying production ML.
Our analysis reveals the characteristics, components, 
and topologies of typical industry-strength ML pipelines
at various granularities.
Along the way, we introduce a specialized data model for
representing and reasoning about repeatedly
run components in these ML pipelines,
which we call model graphlets. 
We identify several rich opportunities
for optimization, leveraging traditional data management ideas.
We show how targeting even one of these opportunities,
i.e., identifying and pruning wasted computation
that does not translate to model deployment, can reduce wasted
computation cost by 50\%  
without compromising the model deployment cadence.
\end{abstract}

\maketitle

\section{Introduction}
\label{sec:intro}

ML (Machine Learning) is now ubiquitous in data-driven organizations, consuming 
an increasing share of compute resources.
A commonly held misconception is that
ML is a one-step training procedure
that accepts data as input and generates
a model as output.
Many research efforts, both within the DB
community and in other communities, 
focus on this simplified view
and aim to improve the effectiveness
of ML training, 
e.g., by generating more powerful models
through better training algorithms,
or by reducing its resource footprint
through various software and hardware
optimizations.
This view is reinforced by various
leaderboard-style competitions
popular among practitioners (e.g., Kaggle),
and academics (e.g., ML benchmarks and KDD competitions).

At the same time, there is evidence from 
practitioners~\cite{sculley2015hidden, DBLP:conf/opml/BaylorHKLLMMPTZ19, 256658} that ML deployments in production 
are significantly more complicated. 
Specifically, ML in production 
involves pipelines with many interlocking steps, 
only one of which is training. 
This has spurred on the development of many end-to-end ML systems (e.g., TFX~\cite{DBLP:conf/opml/BaylorHKLLMMPTZ19}, MLFlow~\cite{zaharia2018accelerating}, Microsoft Azure ML~\cite{azureml}, AWS Sagemaker~\cite{sagemaker}) and open-source ML libraries (e.g., MLlib~\cite{mllib}, MetaFlow~\cite{metaflow}, and Scikit-Learn~\cite{sklearn}), all of which provide native support for data pre-processing, data validation, model validation, and model deployment, in addition to modeling training, all within a single environment.

As an example, TFX~\cite{DBLP:conf/opml/BaylorHKLLMMPTZ19} includes 
pipeline steps that perform 
different flavors of data analysis and transformation, or of data- and model-validation, both before and after the training step. 
The topology of the corresponding
graph can be complicated. 
For instance, model chaining (where a model is used to generate data for another model) is becoming increasingly common, introducing model-to-model dependencies in the same pipeline. And of course, there is the step of deploying a model after training, in a scalable serving infrastructure.
Moreover, production ML pipelines often work in a continuous mode, 
with periodic retraining and deployment as fresh data becomes available.
Overall, the steps across these pipelines interact in complex ways,  
and their compound effects might be hard to predict or debug, necessitating the management of provenance across them. Provenance management is one of the key value adds of existing end-to-end ML platforms, such as TFX~\cite{baylor2017tfx}, MLFlow~\cite{zaharia2018accelerating} or MetaFlow~\cite{metaflow}.  

While there is anecdotal evidence for these
end-to-end concerns beyond training,  
little is known about 
production ML deployments and the challenges they surface 
in terms of data management:

\squishlist

\item {\bf Coarse-grained pipeline characteristics.}
What do production ML pipelines look like
in a large data-driven organization? 
What types of models are used? 
What types of feature engineering procedures are used,
and how complex are they from a data processing standpoint? 
What is the lifespan of typical pipelines? 

\item {\bf Fine-grained pipeline characteristics.}
How much overlap exist between executions of a given pipeline?
For portions of a pipeline that are executed repeatedly to derive models, how does the data distribution change?
How often are they executed, 
and how often are the resulting models deployed?

\item {\bf Opportunities.}
Are there uniform ways to represent and reason about these pipelines?
Any opportunities to make these pipelines more efficient, e.g., 
by leveraging sharing of computation, pruning redundant computation, making more efficient use of system resources, and leveraging incremental view maintenance?

\squishend

\noindent Answering these questions can improve our understanding of production ML. In turn, this increased awareness can help the academic DB community move 
beyond 
training efficiency
to more effectively 
supporting the end-to-end production ML pipelines. 
This involves addressing new incarnations of 
familiar DB challenges---from efficient data preparation and cleaning, to optimized
query plans, to dealing with streaming data, 
to sharing of computation,
to materialization and reuse, 
to provenance for reproducibility and debugging.

In this paper, we take a first step in this direction 
by {\em analyzing a large corpus of 3000
production ML pipelines at Google, comprising over 450,000  trained models,
over a period of four months.}
To the best of our knowledge, no similar
corpus has ever been analyzed in prior literature. 
This unique corpus of  
thousands of TensorFlow Extended (TFX) pipelines, with hundreds of thousands of generated models, spans
different modalities (tabular data, video and 
text embeddings, personalization), 
tasks (regression, classification), 
and environments (production, development). 
Our analysis reveals a number of interesting
insights, including the fact that:
{\em (a)} training accounts for only 20\% of the total computation time, despite $\sim60\%$ of  models being deep neural nets (DNNs);
{\em (b)} the rest 40\% of the pipelines train traditional, non-DNN, ML models, showing the value of simpler model architectures in production, but also the need to manage a diverse set of model types within the same organization;
{\em (c)} the input data used for consecutive model updates have large overlaps but also significant differences in data distribution, underlining the need to cope with data and concept drift,
{\em (d)} models in each pipeline are updated 7 times per day on average, with a substantial fraction (1.12\%) of pipelines updating models over 100 times a day(!), giving rise to potential instability that require special care,
{\em (e)} only one in four model retraining results in model deployment, with the three undeployed model updates representing wasted computation, discussed more later.
While our results stem from analyzing production ML pipelines at Google, due to the commonalities between TFX and other end-to-end ML platforms, as well as the adoption of TFX in other large organizations~\cite{DBLP:conf/kdd/KatsiapisH19}, we expect our findings to generalize to other production use cases and thus be of general interest.
Additionally, we hope our approach to analyzing this complex corpus can serve as inspirations for studying other corpora of ML systems history.

\vspace{2pt}
\noindent Specifically, we make the following contributions:

\topic{Coarse-grained analysis of pipeline lifespan, components, architecture, and complexity (Section~\ref{sec:coarse-grain})} 
We provide the first-ever study of 3000 ML production pipelines captured over a four-month period to understand the underlying data management challenges. This analysis surfaces coarse-grained characteristics about these pipelines, such as their lifespan in the end-to-end training of several models, proportion of resources devoted to data analytics (beyond training), and the features, feature transformations, and model architectures in the pipelines. 
    
\topic{Model graphlet abstraction and fine-grained analysis of frequency, failure, and overlap (Section~\ref{sec:fine-grain})} We proceed to analyze finer-grained properties of these pipelines through provenance analysis. Many previous studies analyzed provenance graphs in data workflows, but the complexity and unique characteristics of production ML necessitate a new approach. We introduce the notion of {\em model graphlets}, wherein the provenance graph is decomposed into sub-graphs to capture the end-to-end execution of the pipeline including several instances of 
model training. One can view model graphlets as a ML-oriented application of the more general concept of provenance segmentation~\cite{DBLP:conf/tapp/AbreuACCEG16}. We then characterize these graphlets in terms of their lifespan, complexity, overlap, failure points, and their connection to model deployment.
    
\topic{Eliminating 50\% of the wasted computation that does not lead to model deployment (Section~\ref{sec:waste})} As an immediate consequence of (and evidence for the value of) our analysis, we identify a significant optimization opportunity involving preemptively skipping pipeline executions.
Overall, there are many wasteful graphlets that neither deploy models to serve applications nor help warmstart subsequent model training.
We show that such graphlets have significant resource costs. Moreover, we show that the root causes for the wasteful graphlets are varied; hence, it is difficult to come up with simple heuristic strategies to identify them. Instead, we leverage the dataset at hand and develop an ML-based solution---we train a model that uses the current state of the pipeline to predict whether the graphlet run will result in a deployed model, without even running it. The model achieves high accuracy and allows us to save up to 50\% 
of wasted computation, without compromising graphlet runs that deploy models. Beyond this direct benefit, this optimization is indicative of the opportunities to optimize ML deployments through a holistic analysis of the ML provenance graph, in addition to localized optimizations of individual steps (e.g., reducing the footprint of the trainer).

\topic{Related Work} We briefly cover previous studies related to tracking, analyzing, and optimizing ML pipelines in production.

\subtopic{End-to-end ML frameworks} Several end-to-end ML frameworks, both commercial and open-source, have been developed recently to address increasingly sophisticated ML use cases. While the programming interface and the runtime may vary across these frameworks, they all aim to support common operators in ML pipelines, including data ingestion and pre-processing, data validation, modeling training and validation, and model deployment.
Systems such as Microsoft Azure ML~\cite{azureml}, TensorFlow Extended (TFX)~\cite{baylor2017tfx} and Kubeflow~\cite{kubeflow}, and open-source libraries such as MLlib (native support in Databricks)~\cite{mllib}, MetaFlow (in production at Netflix)~\cite{metaflow}, and Scikit-Learn (native support in all commercial systems)~\cite{sklearn}, have explicit declarative programming abstractions for composing ML pipelines from canned or custom operators, whereas AWS Sagemaker~\cite{sagemaker} 
captures the notion of an ML pipeline indirectly through integration with libraries with pipeline constructs. 
In terms of the runtime, all commercial cloud-based systems offer capabilities to provision for and schedule ML pipeline executions. 
In addition, TFX and Azure ML also offer automated support for continuous model update and deployment of models.

\subtopic{Provenance management and analysis} Provenance for complex systems has been studied for relational databases \cite{DBLP:journals/ftdb/CheneyCT09}, scientific workflows systems~\cite{DBLP:journals/cse/FreireKSS08}, dataflow systems~\cite{DBLP:journals/pvldb/AmsterdamerDDMST11, DBLP:journals/pvldb/InterlandiSTGYK15}, data lakes~\cite{DBLP:conf/sigmod/HalevyKNOPRW16, hellerstein2017ground}, and ML systems~\cite{zaharia2018accelerating, vartak2016modeldb, DBLP:journals/debu/0001D18}. 
Previous work has even led to the standardization
of provenance representations for workflows in
the form of graphs~\cite{holland2008choosing,DBLP:journals/fgcs/MoreauCFFGGKMMMPSSB11,moreau2013prov}. Other research has proposed various ways to explore and analyze such provenance graphs, e.g., visualization~\cite{DBLP:conf/visualization/BavoilCSVCSF05}, reachability query support~\cite{DBLP:conf/sigmod/BaoDKR10}, support for user-defined views~\cite{DBLP:conf/icde/BitonBDH08}, segmentation and summarization~\cite{DBLP:conf/tapp/AbreuACCEG16,DBLP:conf/cikm/AinyBDDM15,DBLP:conf/icde/0001D19}. 
Our work introduces a framework to segment ML provenance graphs and demonstrates how this segmentation leads to further analysis and optimizations for ML pipelines.

\subtopic{Metadata tracking for the ML Lifecycle} In modern end-to-end ML systems, e.g., TFX~\cite{DBLP:conf/opml/BaylorHKLLMMPTZ19}, MLFlow~\cite{zaharia2018accelerating}, Kubeflow~\cite{kubeflow}, and MetaFlow ~\cite{metaflow}, data science development tools, e.g., ModelDB~\cite{vartak2016modeldb}, noWorkflow~\cite{DBLP:journals/pvldb/PimentelMBF17}, and Pachyderm~\cite{pachyderm}, as well as model development practices in the industry, e.g.,~\cite{schelter2017automatically, DBLP:journals/pvldb/RupprechtDAGB20}, there is an effort towards tracking metadata
to facilitate workflow orchestration and aid model reproducibility over the ML project lifecycle. These metadata tracking solutions vary in terms of ingestion method (user input vs.~transparent), data model (relational vs.~graph), and scope of the metadata (training vs.~end-to-end). Our work here is based on a specific framework (MLMD~\cite{mlmd} which is part of TFX~\cite{baylor2017tfx,blog_tfx_history}) which automatically records the end-to-end provenance using a graph-based model. We build on this framework and introduce a method to segment large provenance graphs into smaller model-centric graphs. However, our analysis and this decomposition is orthogonal from the specific representation of complete provenance, and the same methodology could apply to other systems. Moreover, our final contribution with respect to  pruning redundant graphlets has not been explored in previous work to the best of our knowledge.

\subtopic{Understanding ML workflows}
While our work is, to the best of our knowledge, the first large-scale study of production ML pipelines, a few prior papers have conducted empirical studies on ML and data science (DS) workflows.  
Lee et al.~\cite{hilda} aim to understand how ML developers iterate on models by studying ML workflows generated by novice and intermediate ML developers on Kaggle-style tasks with static datasets and no model deployment. Our study is fundamentally different in the nature of the ML tasks studied---with an emphasis on production ML
and pipelines that are run over a long period. 
Some works in the HCI community study ML/DS workflows by interviewing ML developers and data scientists~\cite{zhang2020data, kandel2012enterprise, amershi2019software}.
Another related area of work is anecdotal reports and retrospectives by industry ML practitioners~\cite{polyzotis2018data, schelter2018challenges,sculley2015hidden} that shed light on real-world ML practices and challenges.
We complement this line of work with 
quantitative insights from a corpus of production ML pipelines.

\begin{figure*}
    \centering
    \begin{subfigure}[b]{0.35\textwidth}
        \centering
        \includegraphics[width=\textwidth]{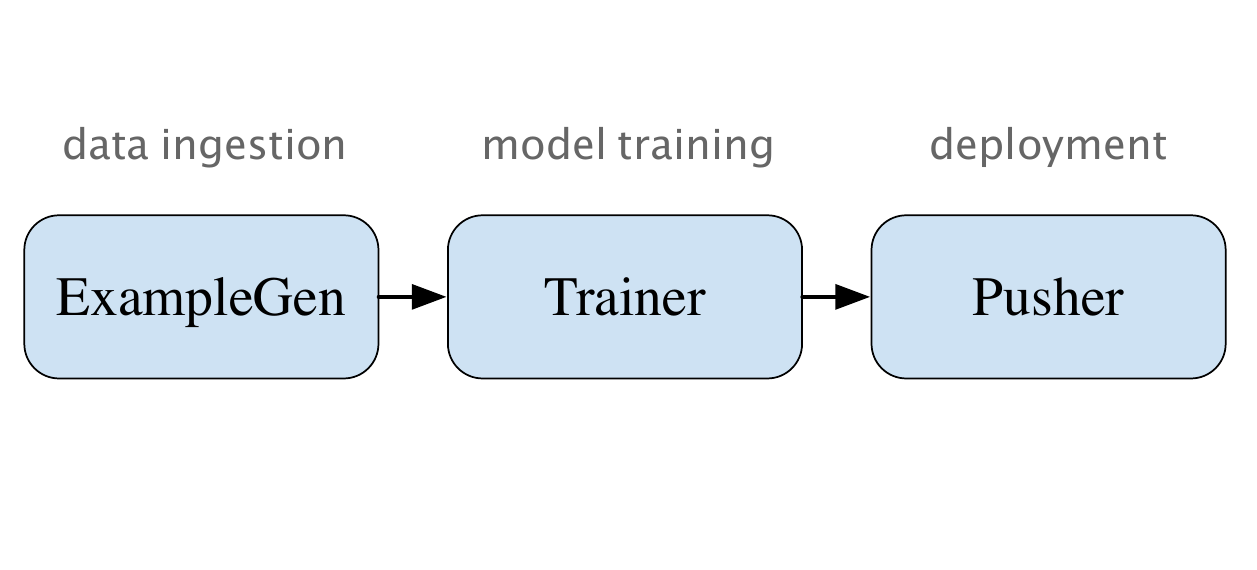}
        \caption{A simple TFX pipeline comprising three operators: \textit{ExampleGen}, \textit{Trainer}, and \textit{Pusher}. Edges denote the input/output dependencies.}
        \label{fig:simple_pipeline}
    \end{subfigure}\hfill
    \begin{subfigure}[b]{0.6\textwidth}
        \centering
        \includegraphics[width=\textwidth]{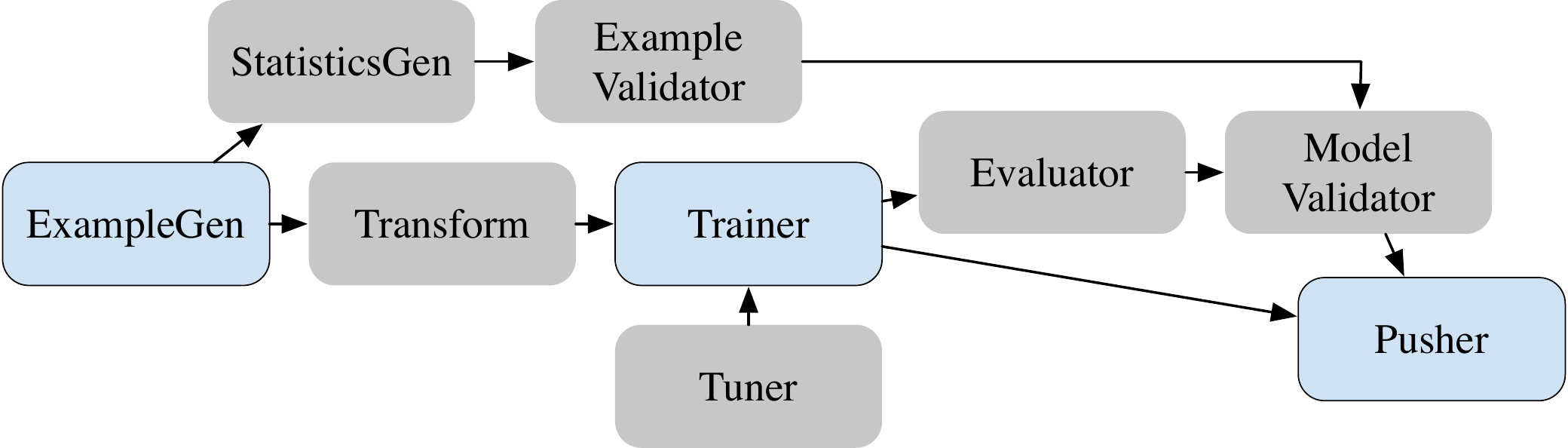}
        \caption{A more typical TFX pipeline that comprises additional operators for data preprocessing, data and model validation, and tuning. Shaded operators correspond to the additional functionality compared to the simple pipeline in \subref{fig:simple_pipeline}.}
        \label{fig:complex_pipeline}
    \end{subfigure}
    \caption{Examples of TFX pipelines}
\end{figure*}

\section{Preliminaries}
\label{sec:preliminaries}

Our corpus comprises TensorFlow Extended (TFX)~\cite{baylor2017tfx} pipelines. TFX is an end-to-end platform for production ML used by product teams across Google. The platform has been recently open-sourced and has been adopted by major organizations outside Google as well~\cite{DBLP:conf/kdd/KatsiapisH19}. TFX provides a set of operators that can be strung together into a pipeline that conceptually accepts data as input and produces a model as output. (The actual topology of the pipeline can get much more complicated in practice, as we describe later.) Even though our discussion here is centered around TFX (to match the actual corpus), we note that the concepts that we introduce are present in other end-to-end ML frameworks discussed previously.

\subsection{Basic Concepts}
\label{sec:basic_concepts}

We use the term \emph{\textbf{pipeline}} to represent a graph of operators that are connected in a producer/consumer fashion. Figure~\ref{fig:simple_pipeline} shows a simple TFX pipeline comprising three operators: \textit{ExampleGen}, which imports data in a format suitable for training; \textit{Trainer}, which uses the imported data to train a model; and \textit{Pusher}, which takes the generated model and deploys it for inference in some external service (e.g., TensorFlow.Serving\footnote{\url{https://github.com/tensorflow/serving}}). 
The topology of the pipeline graph corresponds to the input/output relationships between operators. Moreover, we assume that these relationships are ``type-checked'', e.g., \textit{ExampleGen} outputs data in a format that is understood by the \textit{Trainer}, when the pipeline is authored. The specifics of pipeline authoring are orthogonal to our work.

\begin{figure*}
    \centering
    \begin{subfigure}[b]{0.32\textwidth}
        \centering
        \includegraphics[width=\textwidth]{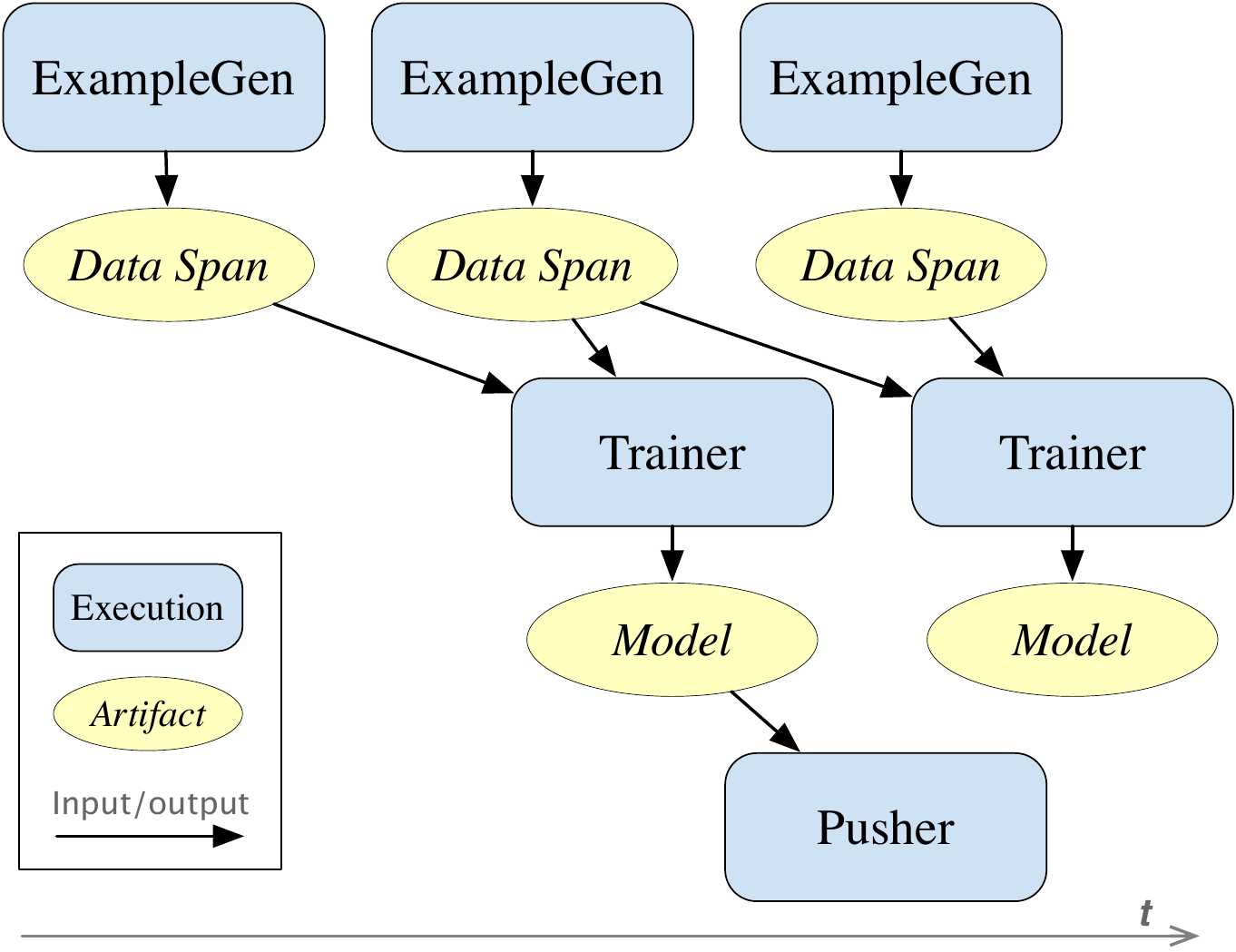}
        \caption{A simple trace for the pipeline in Fig.~\ref{fig:simple_pipeline}.}
        \label{fig:simple_trace}
    \end{subfigure}\hfill
    \begin{subfigure}[b]{0.62\textwidth}
        \centering
        \includegraphics[width=\textwidth]{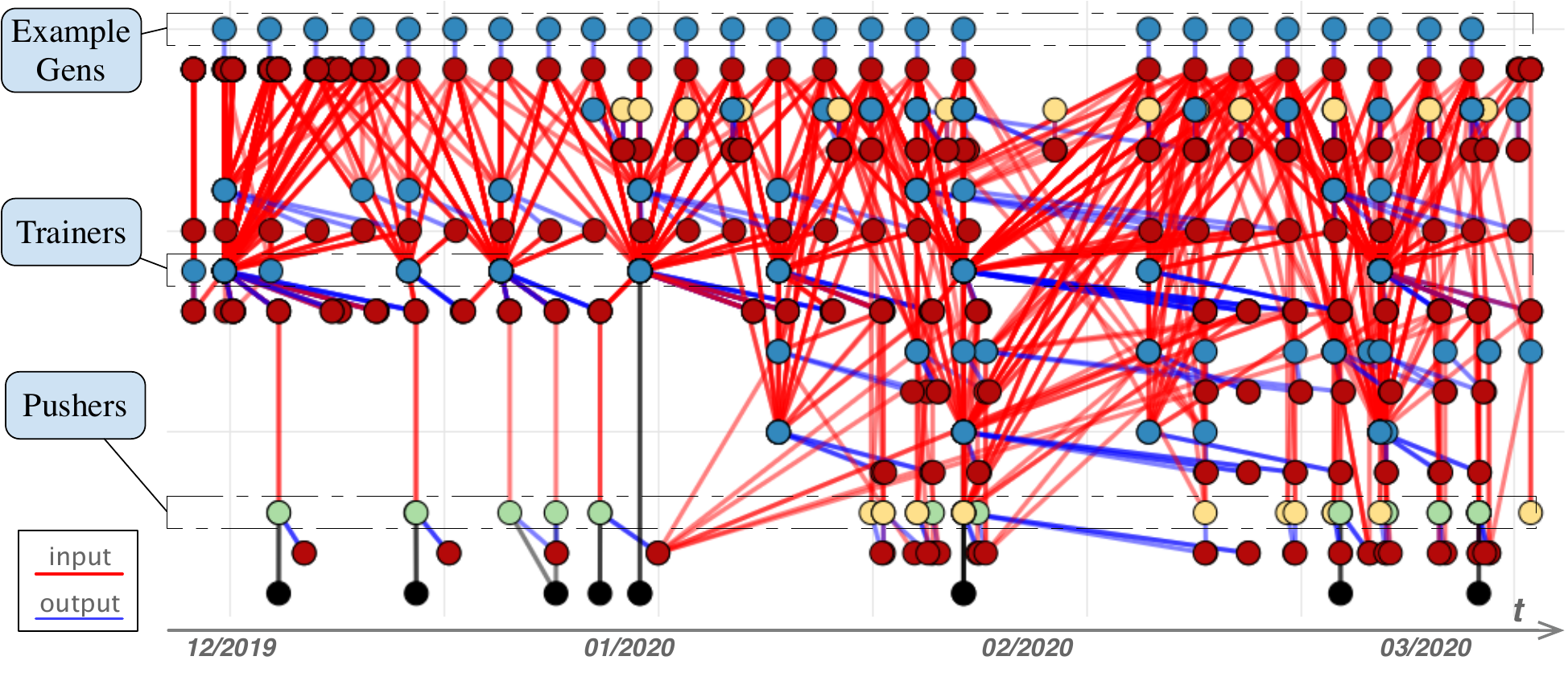}
        \caption{A real-world trace from our corpus, using operators shown in Fig.~\ref{fig:complex_pipeline}.}
        \label{fig:complex_trace}
    \end{subfigure}
    \caption{Examples of pipeline traces.
   The left-to-right order preserves the time of artifact generation.}
    \label{fig:trace_examples}
\end{figure*}

In itself, a pipeline encodes only coarse-grained dependencies between its operators. In contrast, a pipeline \emph{\textbf{trace}} records the fine-grained relationship between individual executions of the operators and the provenance of their input/output artifacts when the pipeline runs in production. Formally, a trace is a directed acyclic graph of \emph{execution} nodes and \emph{artifact} nodes, with edges linking execution nodes to their input/output artifacts. Figure~\ref{fig:simple_trace} shows a sample trace of the pipeline in Figure~\ref{fig:simple_pipeline},
where the nodes are arranged horizontally, left to right, in increasing order of the finish time for executions and creation time for artifacts.
In this trace, the \textit{ExampleGen} operator has executed three times, producing one \emph{data span} artifact per execution. A \emph{data span} artifact corresponds to a chunk of data whose semantics depend on the pipeline. For instance, if the pipeline trains models to recommend videos to users, then a single data span might correspond to previous user interactions with the video service over the period of a single day. Continuing with the example, the \textit{Trainer} operator has executed two times, each time reading the last two data spans in the pipeline and producing the corresponding model. We note that this pattern is quite common in practice: there is a data ingestion process (here, the \textit{ExampleGen} operator) that produces outputs at a fine level of granularity (e.g., each span corresponds to a single day of data) and the model training process reassembles the data into coarser granularity (e.g., a rolling window of the last two days). The example trace concludes with the \textit{Pusher} operator, which executes once for the first model but not for the second model, which means that the second model was not deployed. The latter case is not uncommon in production and can be attributed to several reasons, e.g., the model does not have better performance than the previous model or fails certain validation checks, or the deployment mechanism throttles model pushes to avoid overloading. 

Note that a trace does not reflect any information about the orchestration of the pipeline operators. 
For instance, the example trace might be generated through a serial execution of operators, or by asynchronous scheduling where several operators with overlapping inputs can run at the same time. Moreover, a pipeline may be triggered periodically (e.g., by ingesting the newest span of data every hour and triggering new runs of the operators) or manually (e.g., a model developer reruns the pipeline after making changes to the input data or training code). 
All we assume is that some external system is responsible for scheduling operators, and the trace will grow over time with every run of the pipeline to contain the full history of operator executions and generated artifacts.

Up to this point, we used a simple example comprising the most elemental steps of an ML production pipeline: data ingestion, training, and deployment. In practice, pipelines are more complicated. First, there are several other operators that correspond to important stages and safety checks in production ML. Figure~\ref{fig:complex_pipeline} shows a more typical pipeline that expands on the simple pipeline of Figure~\ref{fig:simple_pipeline} by including operators for data analysis and validation, data pre-processing, and model analysis and validation. Pipeline authors may also introduce custom operators depending on their ML task. Second, these operators can be wired in different topologies, e.g., a Trainer operator might generate an initial model that is then distilled through a separate Trainer operator to produce the final model, or the data-validation operator might block the execution of downstream operators if the data contains any errors. Last, the configuration of the pipeline may change over time. For instance, the pipeline might start with training a single ``production'' model, then be augmented over time with more Trainer operators that correspond to ``experiment'' models (some of which might become production models eventually).

This pipeline-level complexity carries over to the recorded traces. Moreover, it is common to have long-running production pipelines that continuously ingest data spans and output ``fresh'' models, resulting in traces that continuously grow over time. Add to that the fact that executions often share artifacts (e.g., through the rolling-window mechanism described in Figure~\ref{fig:simple_trace}), and it becomes clear that traces can easily become large graphs with complicated structure that are challenging to analyze. Figure~\ref{fig:complex_trace} shows one such large real-world trace; our corpus has pipelines with as many as \textbf{6900 nodes}. This observation motivates our proposal to analyze traces through lower-complexity or finer-grained \emph{model graphlets} that essentially ``unnest'' each trace graph with respect 
to the generated models---more on this in Section~\ref{sec:fine-grain}. 

\subsection{Corpus of Traces for Analysis}
\label{sec:corpus}

The main contribution of this paper is the first-ever analysis of a corpus of production ML pipelines. The corpus comprises the traces of 3000 TFX pipelines (each with up to 6900 nodes) deployed at Google over a period of four months. We focused on pipelines that generated at least one trained model and had at least one model deployed outside the pipeline. 
These are the production pipelines whose models support downstream applications. The resulting pipelines correspond to hundreds of teams that span product areas (e.g., advertising, video recommendations, app recommendations, and maps), ML tasks (e.g., single-/multi-label classification, regression, and ranking), and model architectures (linear and deep models of varying complexity). In total, the collected traces contain 7.7M execution nodes and 20M artifact nodes.

The traces were collected using the ML Metadata~\cite{mlmd} framework (MLMD), which powers the management of metadata and provenance in TFX~\cite{DBLP:conf/opml/BaylorHKLLMMPTZ19}. Similar to our earlier definition, MLMD models a pipeline trace as a graph of \textit{Execution} and \textit{Artifact} nodes with their input/output relationships. We note that the recorded MLMD traces also carry \textit{Context} nodes to represent the grouping of \textit{Artifact}s and \textit{Execution}s, but this information is not used in our analysis. MLMD records additional metadata per node (e.g., start and completion time of \textit{Execution}s, creation time of \textit{Artifact}s), which we use to glean information about the triggering cadence of pipelines. Moreover, the corpus records high-level information for each data span artifact, comprising the features present in the span, their types, and statistics for different feature types (e.g., the unique values for a categorical feature, or the mean and standard deviation for a numerical feature). It's worth mentioning that both TFX and MLMD are open sourced and similar traces with those operators can be derived by other TFX users.

\section{Pipeline-level Analysis}
\label{sec:coarse-grain}

We begin with an analysis of the corpus at the level of entire traces, aiming to understand higher-level characteristics of ML pipe\-lines, such as the structural properties of the corresponding graphs, common data types and transformations, common ML model architectures, and the cost of different operators in addition to training. Section~\ref{sec:fine-grain} will present a finer-grained analysis at the level of sub-traces, to understand the behavior of ML pipelines from the perspective of individual models.

\subsection{Pipeline Lifespan and Activity}

\begin{figure*}
    \centering
    \begin{subfigure}[b]{0.32\textwidth}
        \centering
        \includegraphics[width=\textwidth]{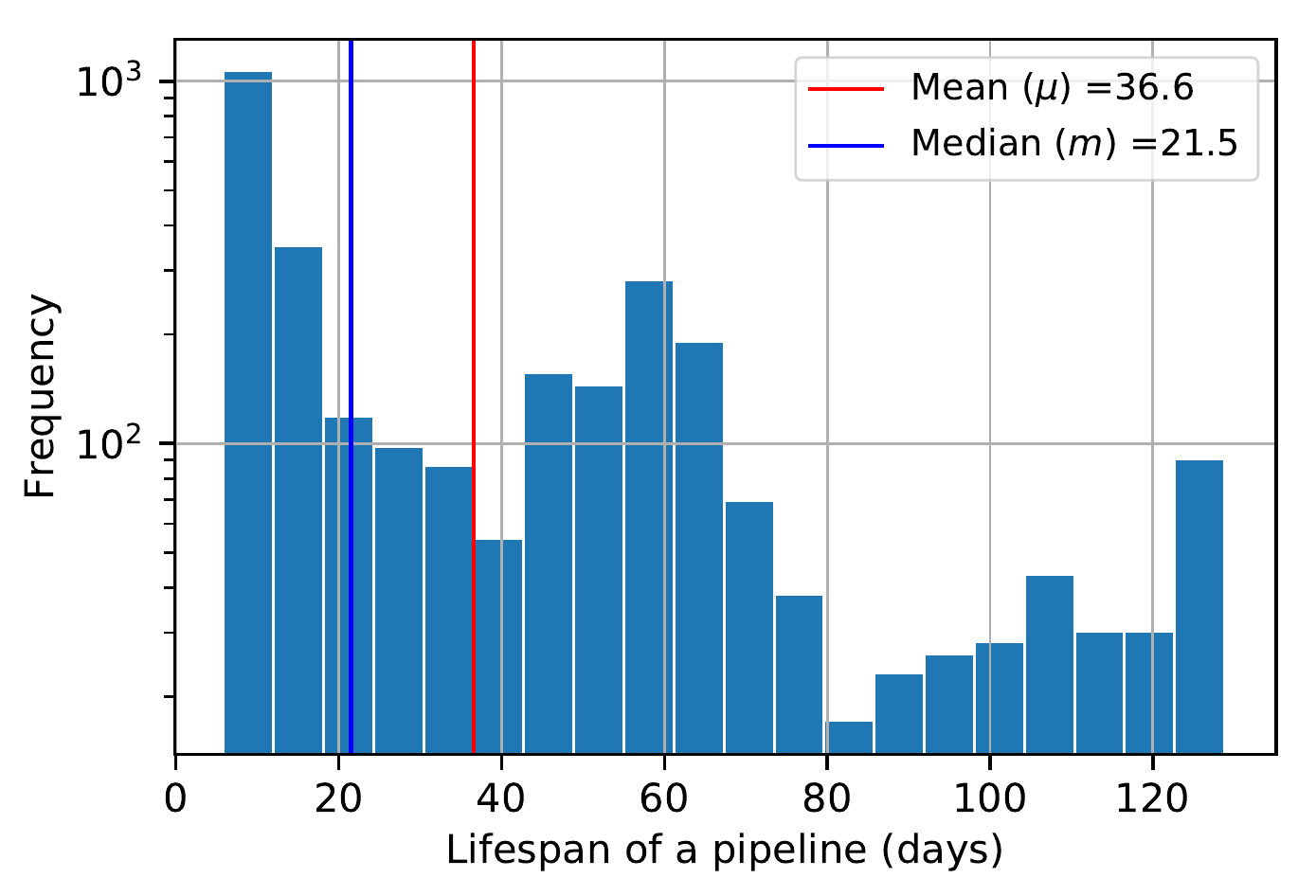}
        \caption{Distribution of pipeline span.}
        \label{fig:lifespan}
    \end{subfigure}\hfill    
    \begin{subfigure}[b]{0.32\textwidth}
        \centering
        \includegraphics[width=\textwidth]{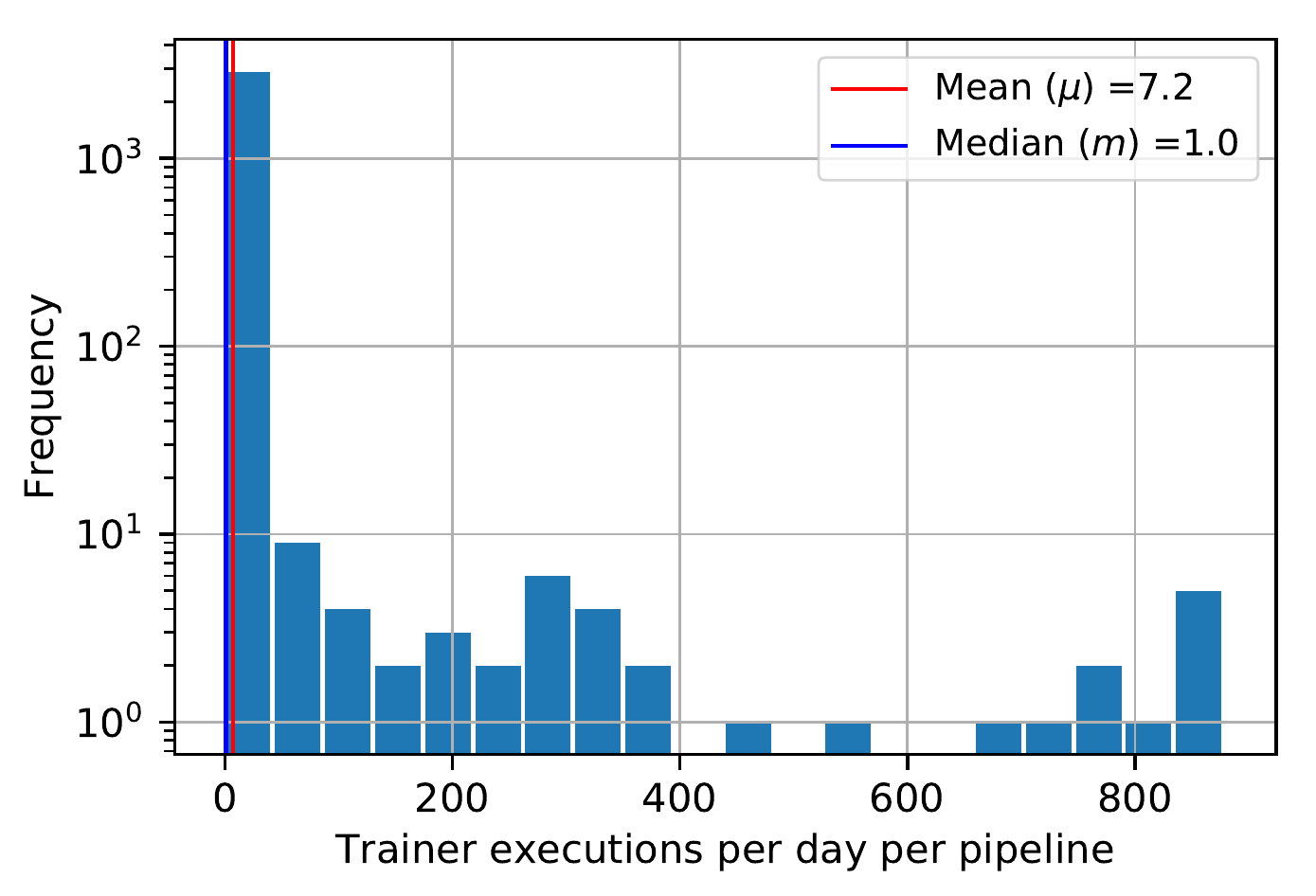}
        \caption{Distribution of trained models per day.}
        \label{fig:retrain}
    \end{subfigure}\hfill
    \begin{subfigure}[b]{0.32\textwidth}
        \centering
        \includegraphics[width=\textwidth]{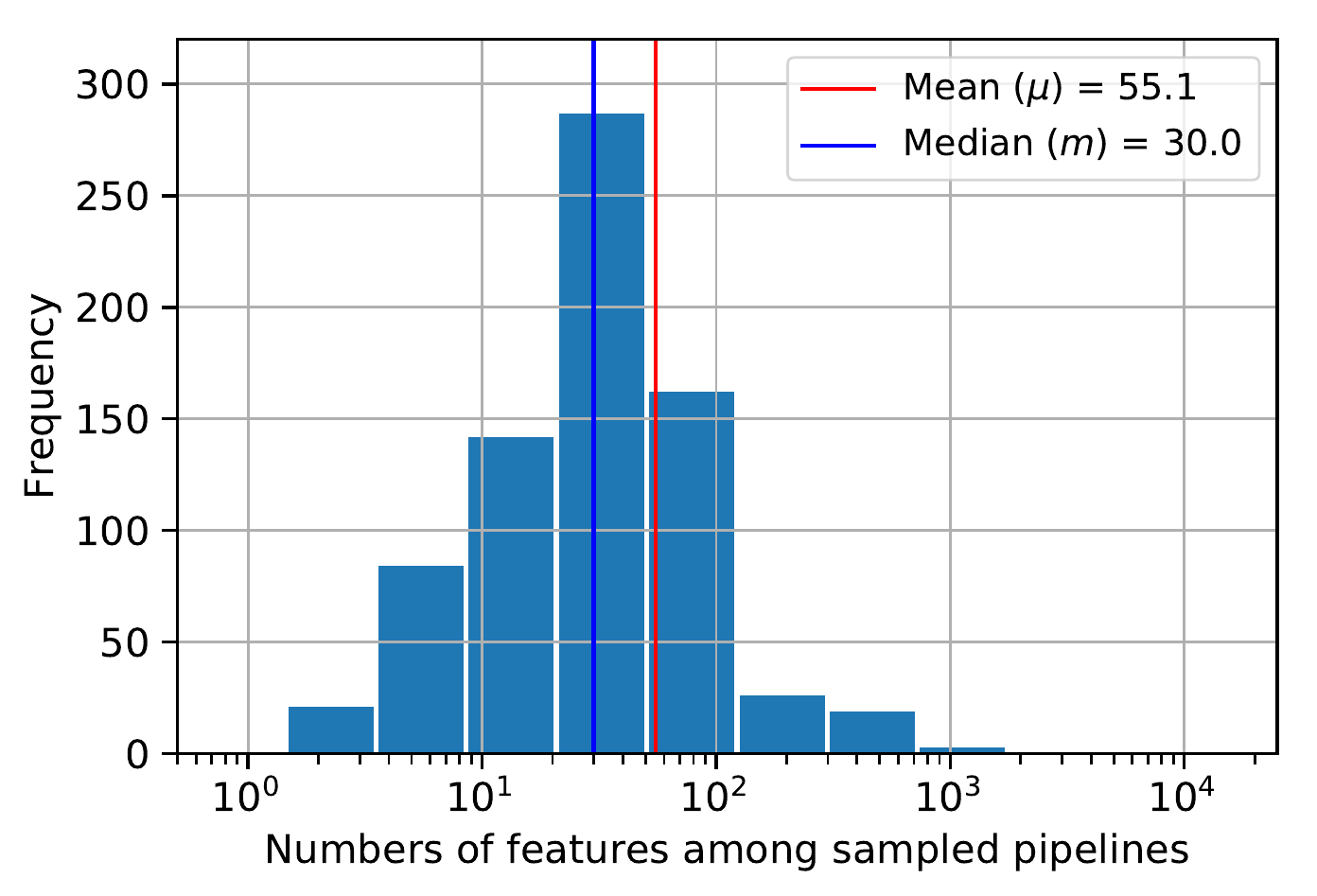}
        \caption{Distribution of the number of features.}
        \label{fig:features_stacked}
    \end{subfigure}\\
    \begin{subfigure}[b]{0.32\textwidth}
        \centering
        \includegraphics[width=\textwidth]{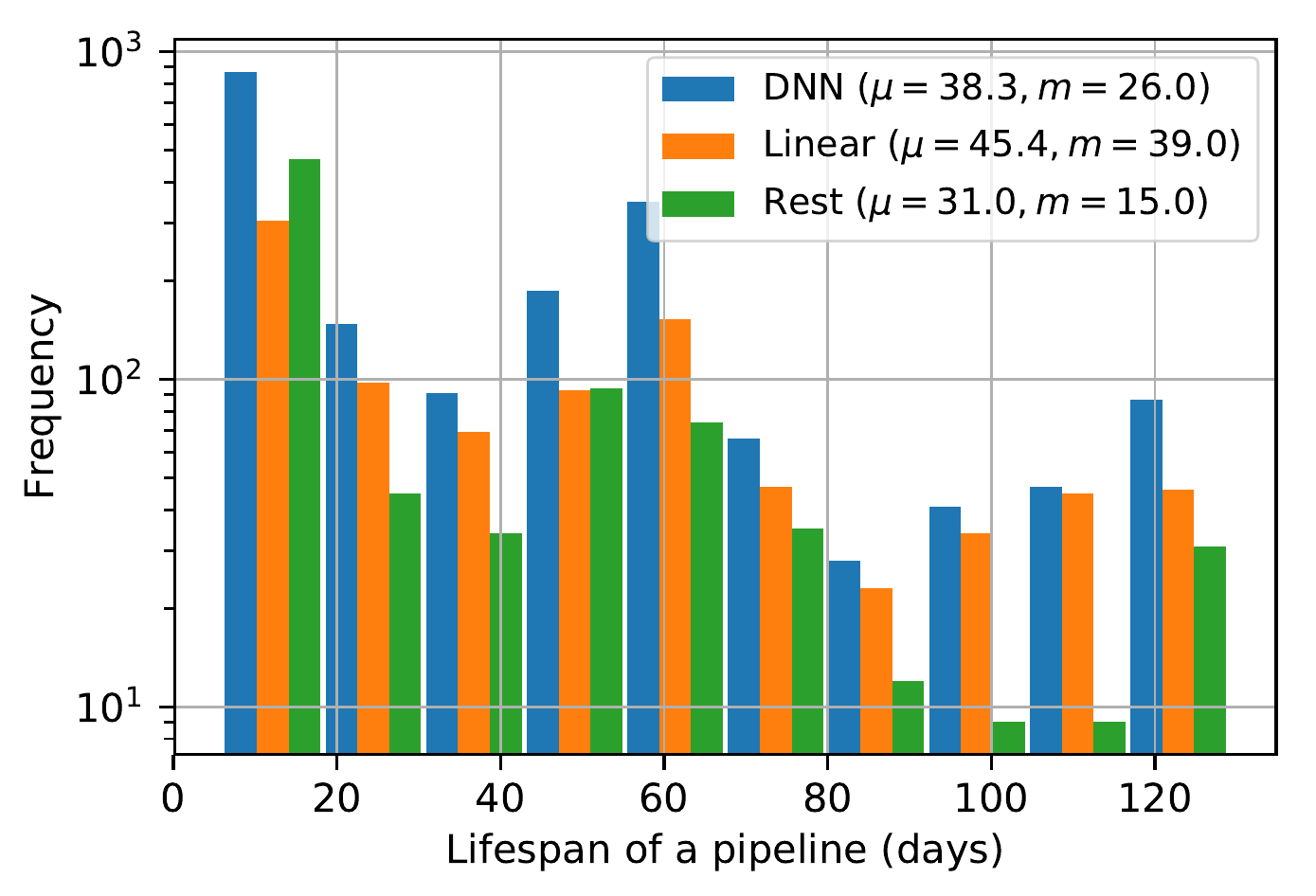}
        \caption{Distribution of pipeline span.}
        \label{fig:lifespan2}
    \end{subfigure}\hfill    
    \begin{subfigure}[b]{0.32\textwidth}
        \centering
        \includegraphics[width=\textwidth]{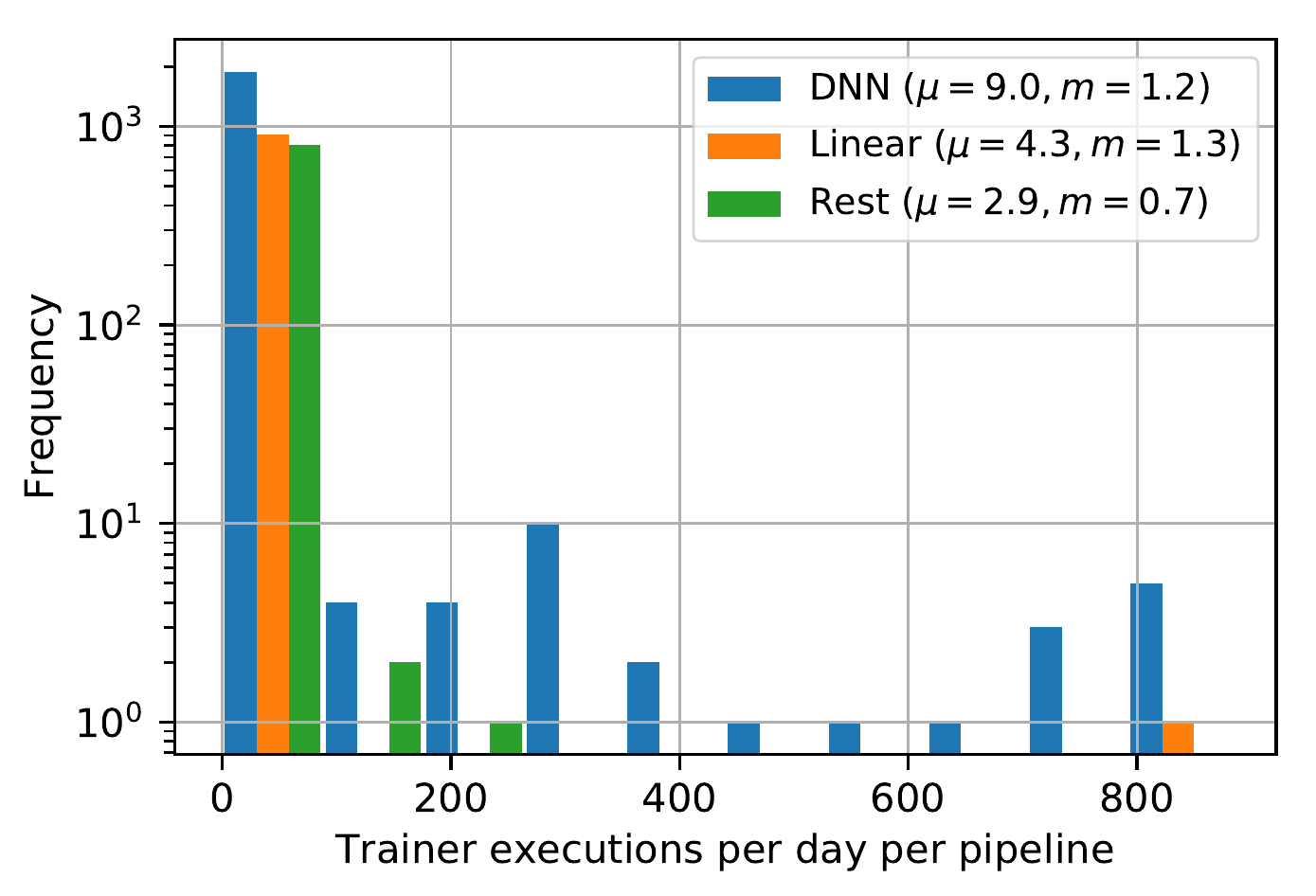}
        \caption{Distribution of trained models per day.}
        \label{fig:retrain2}
    \end{subfigure}\hfill
    \begin{subfigure}[b]{0.32\textwidth}
        \centering
        \includegraphics[width=\textwidth]{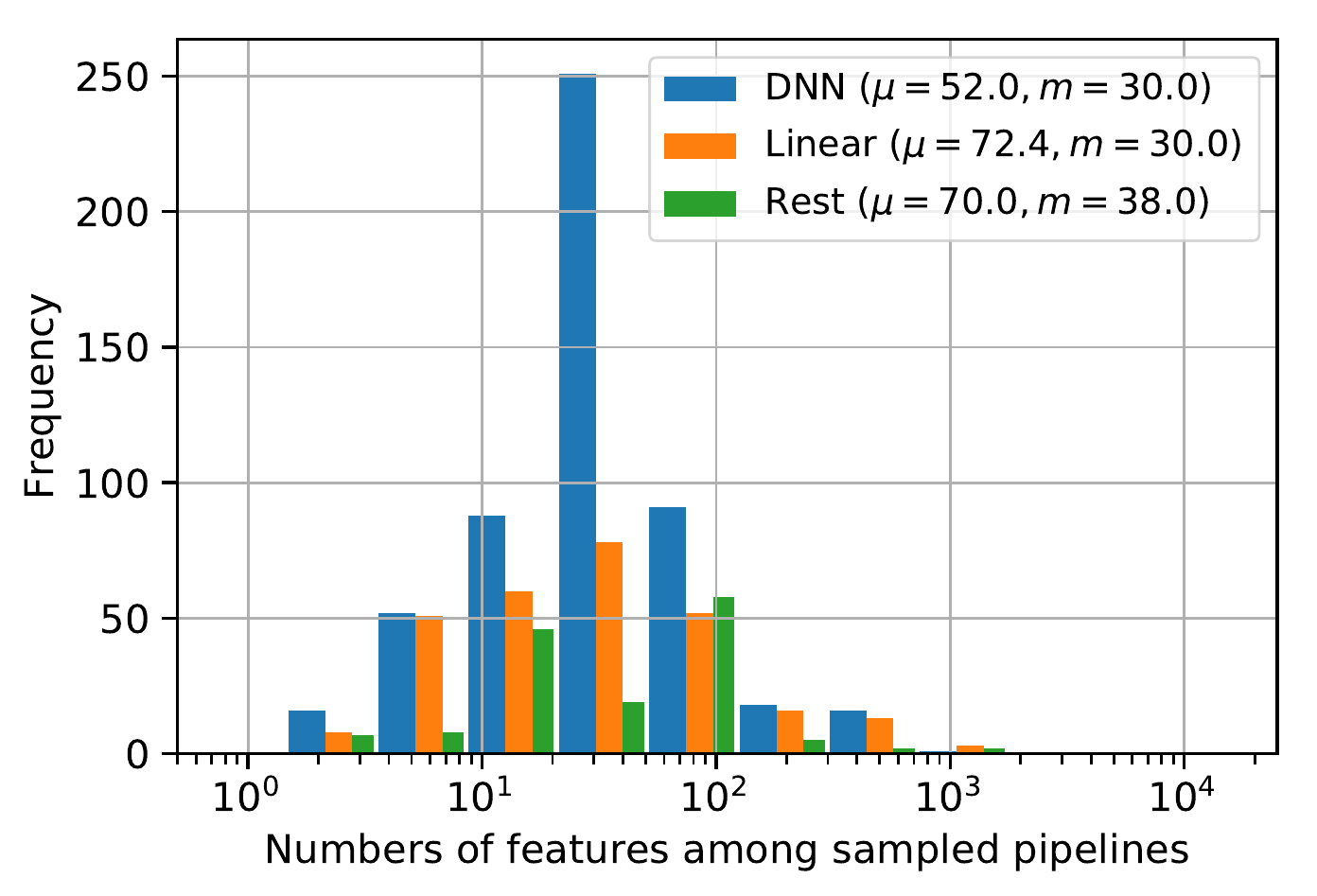}
        \caption{Distribution of the number of features.}
        \label{fig:features_stacked2}
    \end{subfigure}\\
    \caption{Pipeline Activity and Data Complexity Analysis Results.}
\end{figure*}

We first examine two aspects of ML pipelines: 
their typical lifespan 
and model training frequency.We define the lifespan of a pipeline as the count of days between the timestamps of the newest and the oldest nodes
in its trace. Hence, the lifespan is an indication of how long the pipeline is active. The distribution in Figure~\ref{fig:lifespan} shows that, on average, pipelines are active for 36 days, with some being active for the entire span of our corpus (130 days). 
The longer lifespans correspond to continuous pipelines that generate a stream of models based on a stream of incoming data and are common within product groups that are heavy users of ML. Figure~\ref{fig:lifespan2} shows the distribution of lifespan broken down by model type, where ``DNN'' refers to all deep learning models, ``Linear'' refers to all generalized linear models, and ``Rest'' contains all other models, including tree-based models. Interestingly, the pipelines with linear models live longer than the ones with DNN models.

Figure~\ref{fig:retrain} shows a different dimension of pipeline activity in terms of the average number of trained models in the trace, per day. The majority of pipelines generate one model per day but there is also a wide spread of cadences, with some pipelines training close to a thousand models per day! Upon closer investigation, these are continuous pipelines that have a fast stream of incoming data and the ability to quickly produce updated models. As in the previous figure, these pipelines correlate well with product teams that are heavy users of ML. In addition, in Figure~\ref{fig:retrain2}, we find that the cadence of pipelines using DNN are more diverse than the other methods, which reflect a wide adoption of it in different problems.

Overall, pipelines can have a large lifespan and generate models at a high rate. In turn, the accumulated state of these pipelines can become complex. For instance, the trace of the more complicated pipelines can have up to 6953 (artifact or execution) nodes.
Tools to efficiently query or summarize these complex traces can become indispensable for humans to debug or manage these pipelines.

\subsection{Pipeline Complexity}\label{sec:data_complexity}
Next, we analyze the complexity of ML pipelines. Specifically, we examine three aspects: 1) the shape of the input data; 2) the typical transformations for training; 3) the diversity of model architectures.

\begin{figure}[bp!]
    \centering
    \includegraphics[width=0.45\textwidth]{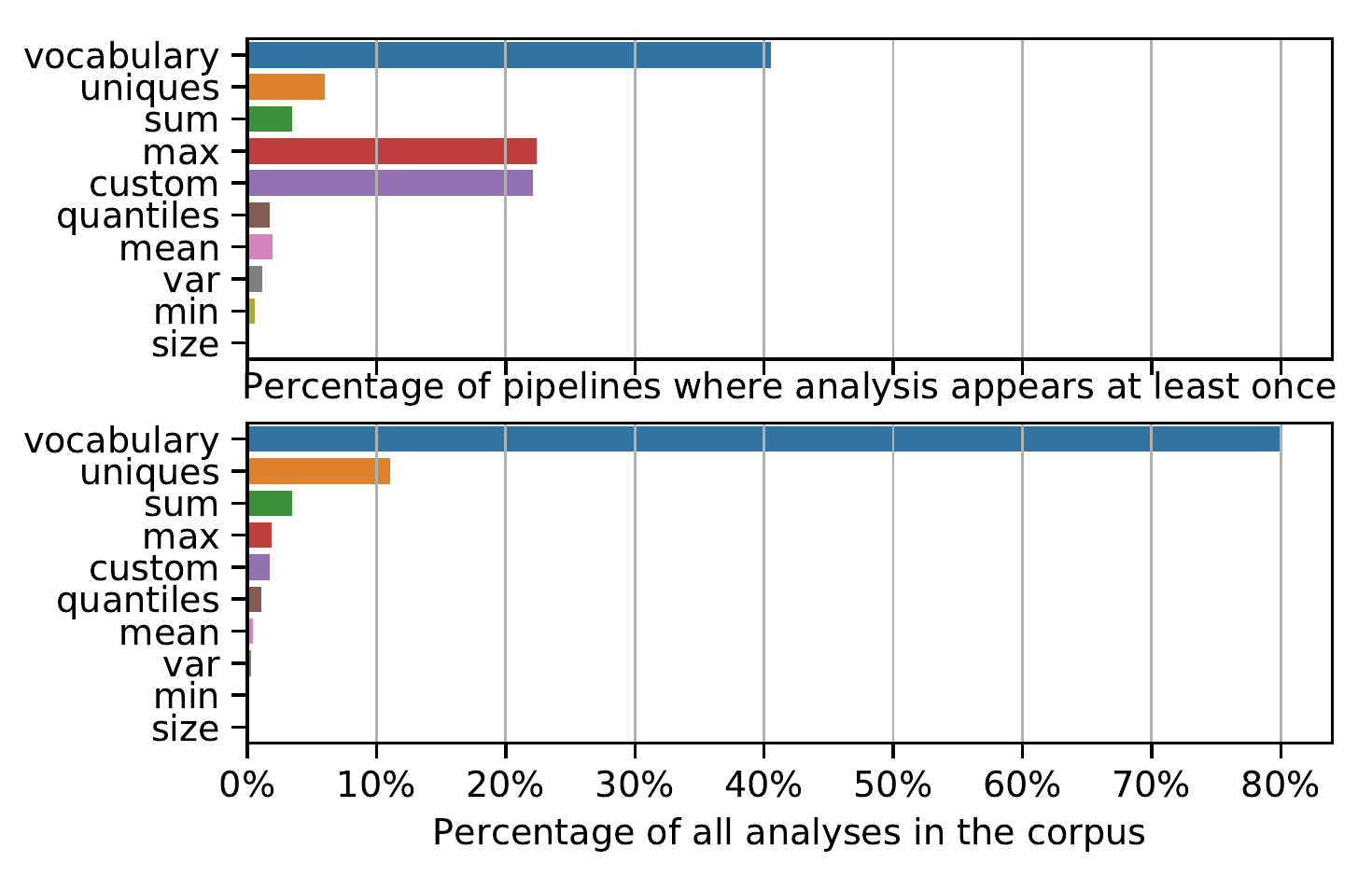}
    \caption{Analyzer Usage}
    \label{fig:analyzer_sum_and_count}
\end{figure}

\topic{Input Data Shape} Figure~\ref{fig:features_stacked} and Figure~\ref{fig:features_stacked2} show the distributions of the feature count in the input data. We use ``feature'' to refer to a column in the data (e.g., the `video-id' column) and ``domain'' for the set of values that this feature has in the input examples. We compute the distribution of feature count as follows: we identify the data span nodes in all traces, and use the associated MLMD metadata for each data span (Section~\ref{sec:corpus}) to retrieve the number of features. As shown, the vast majority of the pipelines utilize up to 100 features. However, higher feature counts do appear and there are extreme cases of tens of thousands of features in a pipeline. 

We also examine the composition of features in each pipeline in terms of numerical (e.g., length of a video) and categorical/sparse features (e.g., id of a video, query text). Note that this distinction does not necessarily reflect the type used to encode the feature. For instance, a `video id' feature might be encoded as a number but treated as a sparse/categorical feature in training as well as in our breakdown. We find that the features of each pipeline are equally distributed among these two types: on average, 53\% of the features are categorical. Moreover, we analyzed the domain of the categorical features and found that each categorical feature has, on average, \textbf{10.6 million} unique values in its domain. For pipeline with DNN models, the average is 13.6 millions, while for the ones having Linear models, it is larger than 20 million. This large domain size indicates high data complexity and thus increased cost and complexity for data transformations prior to training. For instance, we discuss below how such categorical features are typically embedded in vocabularies prior to training.

\topic{Feature Transformation} The additional metadata that we collect in our corpus (see Section~\ref{sec:corpus}) provides a glimpse into the types of transformations applied to the raw data before model training. These transformations are often applied in two stages: the (optional) first stage performs an analysis of the data to derive required statistics for the transformations deriving necessary statistics for the transformation; the second stage uses these statistics to apply the transformations. As an example, consider the common z-score transform for numerical features: the first stage computes the mean and standard deviation of the feature values, and the second stage uses these two metrics to normalize each feature value. In general, the second stage is embarrassingly parallel and can thus easily scale to large datasets. The first analysis stage, however, is much more expensive as it requires potentially expensive reductions over the data (e.g., sorting a large space of video ids based on frequency). We thus focus on this analysis stage in what follows, since that is where we see opportunities for data processing-related optimizations. 

Figure~\ref{fig:analyzer_sum_and_count} shows the different types of analyses applied in the first stage of feature transformations. The ``vocabulary'' analyzer performs the aforementioned truncation of a sparse feature into a smaller numerical domain. As an example, given a feature that contains text tokens (e.g., words), this analysis computes the top-K tokens based on frequency and then maps the features to the numerical domain $[0, K]$. The other types correspond to more straightforward analyses over numerical features (e.g., min, max, and so on). Finally, ``custom'' refers to a black-box analysis (essentially, a UDF) that is pipeline-specific and tailored to the business logic of the corresponding ML task. Pipeline authors resort to these UDF-style analyses when their features require more complex handling than what TFX offers out of the box.

Figure~\ref{fig:analyzer_sum_and_count} shows two views of the same data. 
At the top, we show the percentage of pipelines that reference each analysis at least once, which indicates the relevance of these analyses across 
the pipelines in our corpus. The bottom view shows the total usage of these analyses across all traces and indicates the frequency of their usage in production. Both views confirm the prominence of 
vocabulary computations over categorical features. This becomes even more pronounced when looking at the actual usage in traces (bottom view). We observe 
that custom analyses are also used in several pipelines, although their total usage is much lower in the bottom chart. Our hypothesis is that such UDF-based 
analyses are more relevant for experimental pipelines that have a short lifespan and less activity, whereas canonical analyses provide good coverage for 
``steady-state'' pipelines. 

Both views in Figure~\ref{fig:analyzer_sum_and_count} confirm the prominence of vocabulary computations over categorical features, which provides an interesting opportunity for the data management community, in two respects. First, this computation is a top-$K$ query over an aggregation of the data where $K$ can be very large, e.g., values of $K$ from hundreds of thousands to millions are not uncommon. It is interesting to consider how these queries can be optimized for different representations of the data. Second, the choice of $K$ has non-trivial implications on model quality and performance. A higher $K$ implies better coverage of ``important'' sparse values, which in some cases results in significant improvements in accuracy. At the same time, since this mapping becomes part of the model, a higher $K$ implies larger model sizes (to store the mapping) and perhaps higher processing time (to perform the mapping). To make an informed choice, the model developer needs to understand the tradeoffs for different values of $K$ in terms of these dimensions, and this, in turn, introduces an interesting data analysis problem that may be amenable to techniques from data management (e.g., approximate query answering for this class of queries). 

\begin{figure}[tbp!]
    \centering
    \includegraphics[width=0.45\textwidth]{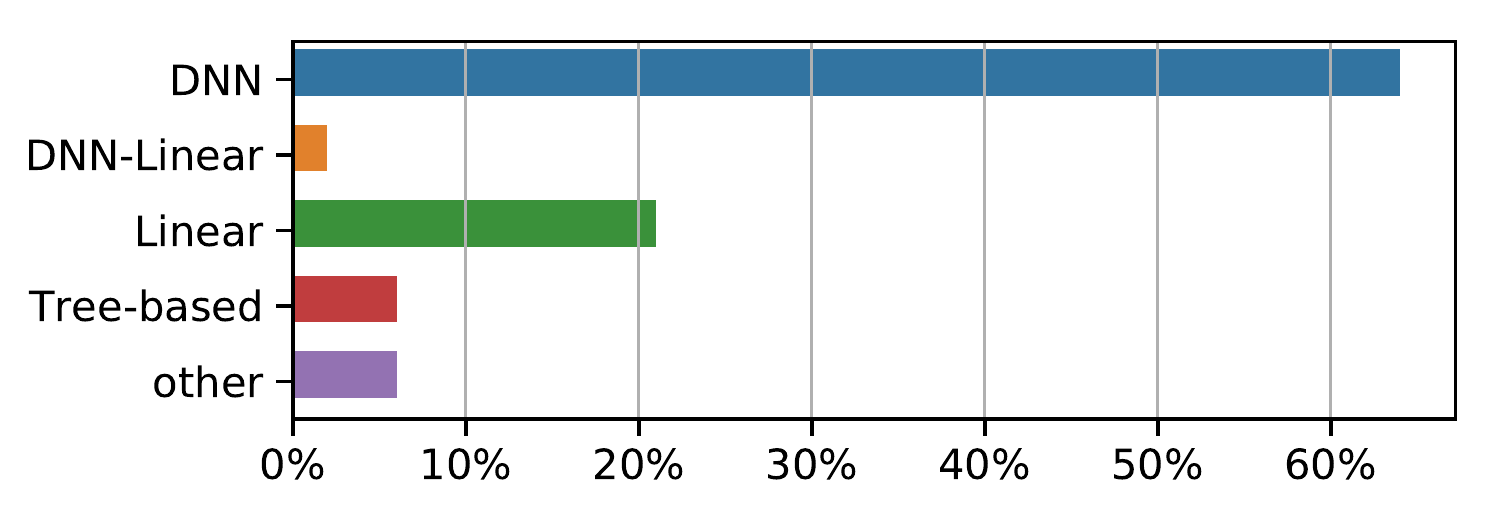}
    \caption{Percentage of Trainer runs with each model type}
    \label{fig:models}
\end{figure}

\topic{Model Diversity}
We next examine the type of models used across pipelines. Doing so helps us characterize the diversity of training methods in production. Moreover, the choice of model architecture affects the selection of other operators and hence other characteristics of the pipeline. 
Figure~\ref{fig:models} shows the usage of different architectures as a percentage of all models in our corpus. 
As we can see, 64\% of the Trainer runs use deep neural networks (DNNs), with an additional 2\% that use a combination of DNNs with linear models. A smaller percentage of pipelines employ linear models and tree-based methods. Finally, a small fraction of the pipelines deal with specialized tasks requiring ensemble models or custom methods that we lump in the ``other'' slice. 

It is interesting to relate this breakdown to recent papers that embed ML techniques inside database management systems and thus aim to jointly optimize data access and model training~\cite{DBLP:journals/sigmod/KumarMNP15, DBLP:journals/pvldb/HellersteinRSWFGNWFLK12, DBLP:journals/pvldb/Olteanu20, karanasos2019extending}. A focus on a specific class of models (e.g., linear or DNN models) is still relevant in practice and can cover a significant fraction of production ML workloads. However, this would leave on the table a much bigger fraction of pipelines outside of the selected class that could also benefit from optimized systems. Moreover, the diversity shown in Figure~\ref{fig:models} is an indication that ML practitioners need access to a wide range of choices, and hence are likely to ``outgrow'' a system that offers a few choices for model architectures.

\subsection{Resource Consumption}

Finally, we turn our attention to the composition of pipelines in terms of
the different operators and their corresponding resource footprints. 

\begin{figure}
    \centering
    \includegraphics[width=0.48\textwidth]{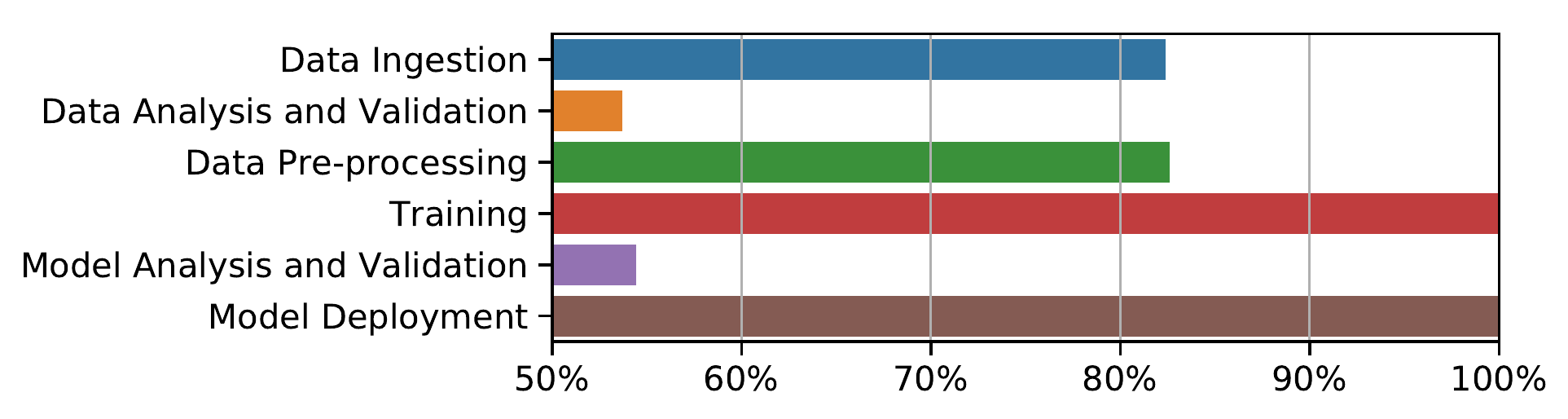}
    \caption{Percentage of pipelines with different operators.}
    \label{fig:component_binary}
\end{figure}

Figure~\ref{fig:component_binary} shows the different types of TFX operators present in our traces and the corresponding percentage of pipelines using these operators. Here we group them in terms of their high-level functionality in the pipeline: data ingestion; data analysis and validation; data pre-processing;  training; model analysis and validation; and model deployment.
Perhaps unsurprisingly, the most common operators include data ingestion, data pre-processing, training, and deployment, with training and deployment in 100\% of the pipelines since our corpus focuses on ML pipelines that support production applications.
It is also worth noting that about half of the pipelines employ data- and model-validation operators, which essentially block the deployment of the trained models if there are errors in the data or if the model metrics are not sufficiently good, respectively~\cite{DBLP:conf/mlsys/BreckP0WZ19, DBLP:conf/sigmod/DrevesHPPRG20}. These operators act as safety checks and are common in pipelines that trained models used in downstream production systems.  

\begin{figure}
    \centering
    \includegraphics[width=0.48\textwidth]{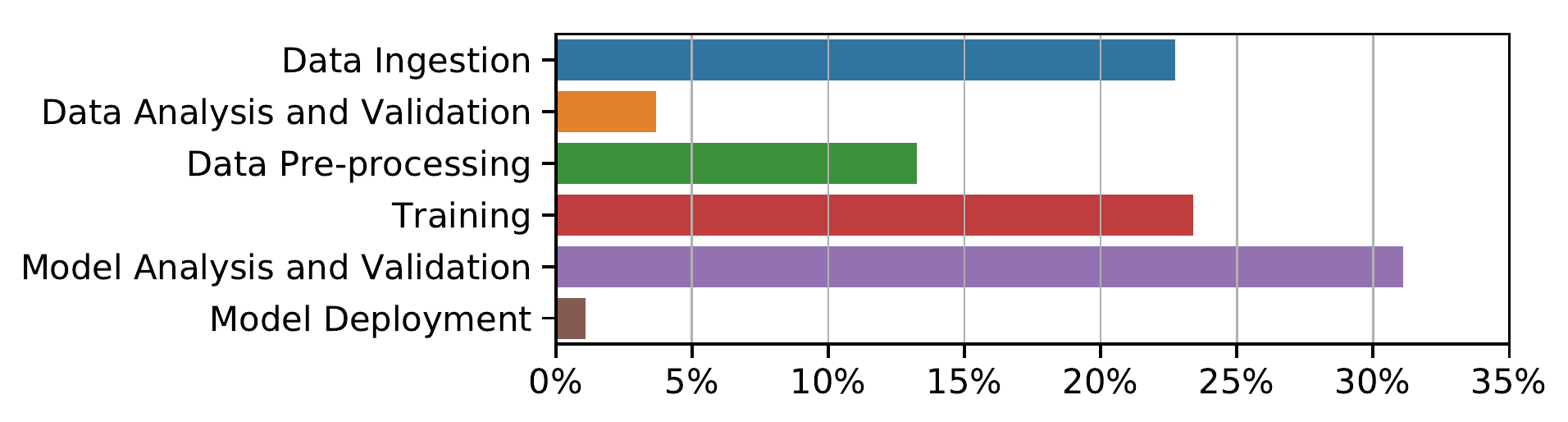}
    \caption{Compute cost of different operators.}
    \label{fig:comp_cpu}
\end{figure}

Figure~\ref{fig:comp_cpu} shows a different breakdown of operators based on their resource usage. For each group, the figure shows the total compute cost as a percentage of the overall compute cost by all groups. 
Perhaps surprisingly, we see that the training operators account for less than a third of the total computation. This finding goes against the conventional picture that ML is mostly about the training algorithm. In contrast, our results indicate that production ML involves more steps that are also more costly, e.g., the data/model analysis and validation operators account for $\sim$35\% of the total compute cost of the pipelines and are more expensive than training. Note that like data analysis and validation, model analysis and validation also boils down
to traditional data processing operations, since it involves the computation of model metrics (e.g., average loss, area-under-the-curve) on slices of the input data, i.e., group-by queries with a model-driven aggregation per group. The figure also shows a significant cost for data ingestion ($\sim$22\%). This happens because, in many cases, TFX initiates a ``hermetic'' copy of each data span from the external data source, along with shuffling/partitioning of the examples to different splits (training/testing/eval) for the downstream model. This cost can be avoided for data sources that provide snapshot-based access with randomization guarantees for the individual examples.

Besides showing the cost of operators beyond training, the breakdown in Figure~\ref{fig:comp_cpu} reveals one more point: pipeline failures can be costly in terms of resources. Specifically, ML pipelines have several failure points corresponding to the different operators, e.g., the pipeline may stop because the data contains errors, or because the training code has faults, or because model validation failed. 
In turn, each failure point may occur after the successful completion of upstream operators, several of which can be costly based on our analysis. In other words, failures are not cheap and there is an upside to preventing them or dealing with them proactively (more on this in Section~\ref{sec:waste}). Moreover, artifact caching and reuse, when feasible, can cut down significantly on the cost of different stages. One example is restarting a failed pipeline due to errors in the training code---since the data has not changed, it should be feasible to reuse any data transformations and thus avoid the cost of re-analyzing the data.
We can use the costs in Figure~\ref{fig:comp_cpu} (in conjunction with failure probabilities)
to determine optimized materialization policies, identifying where it might be most
valuable to cache artifacts, e.g., after pre-processing, training, or model validation, as was done in recent work~\cite{xin12helix}.

\section{Fine-Grained Graphlet Analysis}
\label{sec:fine-grain}
The previous section presented a coarse-grained analysis that focused on aggregated characteristics across pipelines. In this section, we dial up the resolution of the analysis to investigate fine-grained characteristics within each pipeline, and in particular, the cadence and characteristics of model training and deployment---which, as stated earlier, can happen several times within a trace.

\subsection{Model Graphlets}

\begin{figure}
    \centering
    \includegraphics[width=0.4\textwidth]{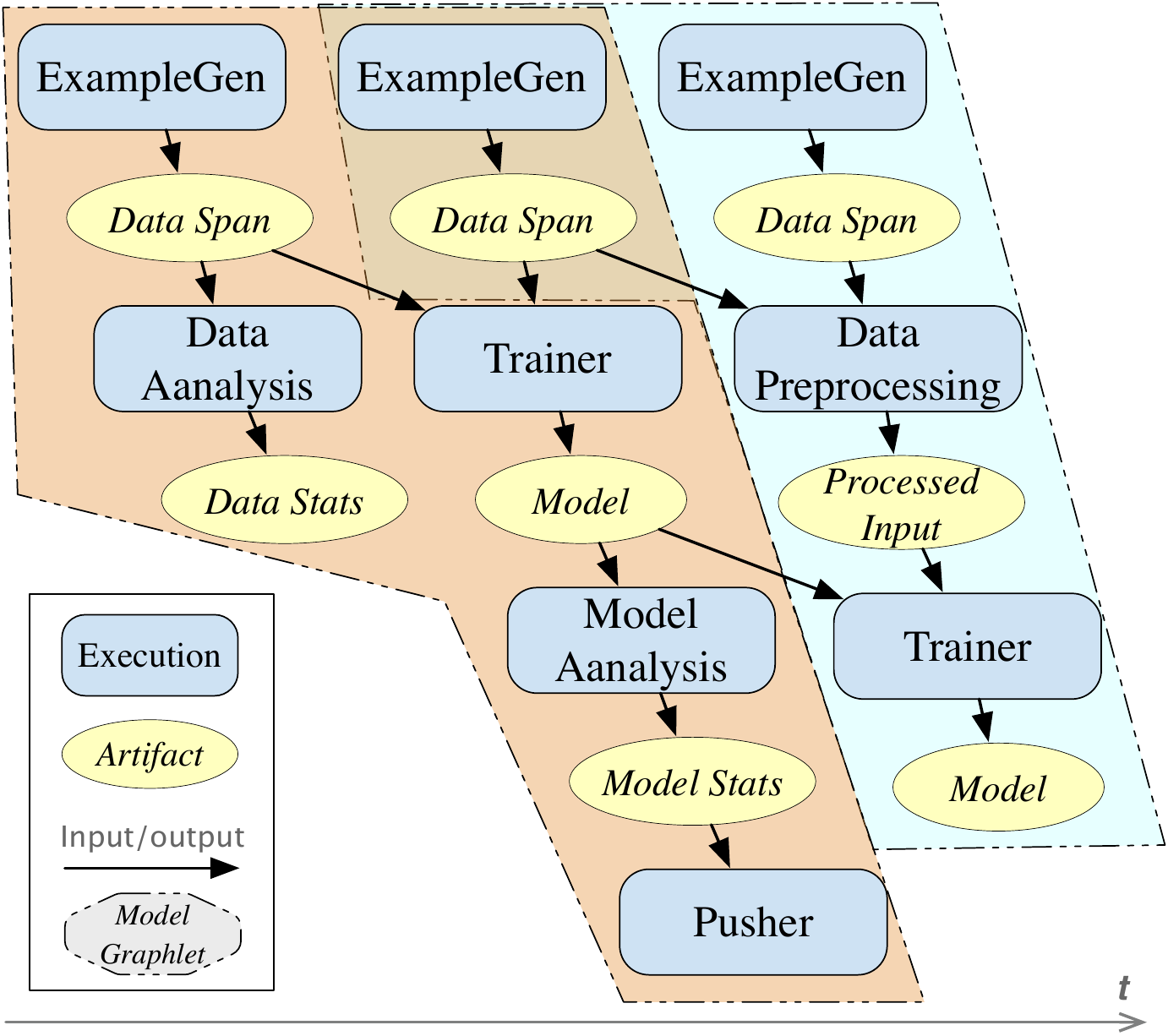}
    \caption{A model graphlet example}
    \label{fig:extraction}
\end{figure}

Recall that pipelines can be continuous, in that they consume a stream of incoming data spans to train and produce fresh models. Hence, a single pipeline trace may contain multiple executions of the Trainer operator (and associated downstream and upstream operators) on overlapping inputs. For instance, when using a rolling-window on the input data spans to update models, a data span might contribute to several models; when using warm-start during training, a generated model artifact might be used as an input to other Trainer executions. As a result, the pipeline trace is intricately tied to the complexity of  ML in practice, and it is not uncommon for the trace to have a single large connected component that encompasses all nodes, which in turn makes it difficult to examine the finer-grained characteristics of the pipeline (e.g., Figure~\ref{fig:trace_examples}).

The issues of inter-connectedness and size of provenance graphs 
have similarly emerged in different domains,
wherein techniques such as user views, segmentation, and aggregation 
have been explored to transform the graphs to usable or interpretable ones~\cite{DBLP:conf/icde/BitonBDH08, DBLP:conf/tapp/AbreuACCEG16, DBLP:conf/icde/0001D19, DBLP:journals/corr/Moreau15}. 
We adopt a similar approach but we leverage the semantics of production-ML operators and connections between them. Specifically,  
we segment the trace graph into several subgraphs, so that each subgraph represents a single (logical) end-to-end pipeline run related to an individual model. 
We provide some intuition for this definition in the context of an example below.
We refer to such a subgraph as a \textit{model graphlet}, or graphlet for short. The segmentation works as follows: given a Trainer execution node $n$, the corresponding graphlet $g$ is a subgraph of
the pipeline trace that comprises
the following nodes, and the corresponding edges: 

\begin{enumerate}[a.]
    \item all ancestor executions of $n$ (and their input/output artifacts);
    \item all data-analysis/-validation executions performed on data spans from the previous step (and their input/output artifacts); and 
    \item all descendant executions of $n$ (and their input/output artifacts) that are not on paths to other Trainer executions. 
\end{enumerate}

\smallskip \noindent
At the highest level, rule {\em a} captures
all ancestor executions that directly influence $n$ along with their artifacts; rule {\em b} recognizes the fact that
there may be data analysis/validation executions performed on data spans that are input to $n$ or one of its ancestor executions,
but are not themselves ancestors of $n$,
and therefore includes any such data analysis/validation executions;
rule {\em c} adds descendent executions 
of $n$, post-training, 
that are not counted in the graphlets of
other Trainer executions---this is checked
by ensuring that they are not on paths
to such Trainers. We provide the Datalog queries for the graphlet segmentation in Appendix~\ref{sec:datalog}.

\topic{Example} Figure~\ref{fig:extraction} shows the graphlets extracted on a sample trace. Intuitively, each graphlet corresponds to a model and captures the subgraph of the trace that is relevant for the generation (rule {\em a}) and deployment (rule {\em c}) of the model. Note that the first model is used to warmstart the Trainer for the second model, yet per rule {\em c} this edge is a ``cut'' between the two graphlets. The reason is that we aim to create smaller graphs of bounded complexity, so that we can analyze the corpus at the granularity of individual models.
The graphlets also include data-analysis artifacts (rule {\em b}), so that we can analyze data-related metrics (re-use and similarity) across graphlets. 

From this point onward, we recast our analysis on the set of graphlets that we extract from the corpus. In total, this gives rise to 450,000 graphlets.

\begin{table}[b!]
    \centering
    \begin{tabular}{c|c|c|c|c||c}
          & [0, 0.25]  & (0.25, 0.5] & (0.5, 0.75] & (0.75, 1] & $\mu$  \\
          \hline
        \textbf{Jaccard} & 30.2\% & 8.2\% & 4.4\% & 57.3\% & 0.647\\
        \hline
        \textbf{Dataset} & 89.7\% & 0.3\% & 0.1\%& 9.9\% & 0.101\\
        \hline
        \textbf{Avg Dataset} & 87.3\% & 5\% & 3.1\% & 4.6\%  & 0.092\\
    \end{tabular}
    \vspace{2mm}
    \caption{Similarity metrics for consecutive model graphlets; percentage of consecutive graphlet pairs in each similarity range, along with the mean similarity.}
    \label{tab:similarity_histograms}
\end{table}

\subsection{Data Change across Graphlets}
Data characteristics,
and the evolution thereof, can have a  significant impact on ML model performance~\cite{polyzotis2017data}. We thus begin our analysis by investigating data evolution across graphlets of the same pipeline. Specifically, we seek to understand to what extent data is reused across different graphlets 
and the rate at which data distribution shifts over time. Answering these questions can provide useful insights related to ML data management, e.g., whether materialization of intermediate data transformations and incremental computation can be useful for ML pipelines.
The analysis that follows uses the notion of \emph{consecutive graphlets}. Two graphlets $g$ and $g'$ from the same pipeline are said to be consecutive iff the corresponding trainer executions are adjacent in chronological order. 

\subsubsection{Reuse}\label{sec:jaccard-sim}
We quantify data reuse with the Jaccard similarity between the data spans of consecutive graphlets $g$ and $g'$, i.e., $|\mathcal{I}(g)\cap\mathcal{I}(g')|/|\mathcal{I}(g)\cup\mathcal{I}(g')|$, 
where $\mathcal{I}(g)$ is the set of input data spans in $g$.

The first row in Table~\ref{tab:similarity_histograms} shows the histogram of the similarity values in the corpus.
The high value at (0.75, 1]---57\% of all pairs of consecutive graphlets---is due to the fact that many pipelines train multiple parallel models on the same inputs for A/B testing. Moreover, consecutive graphlets can correspond to retrainings on the same data after the pipeline author changes other details of the process, e.g., the training algorithm or feature transformations. 
The mean similarity of .647 indicates that, on average, graphlets share two thirds of the data spans. Moreover, several consecutive graphlets have a $>80\%$ similarity of their inputs. These results point to interesting optimization opportunities on data preparation for training (e.g., data pre-processing, data transformation,  or data validation). Specifically, we can leverage this overlap and employ techniques from incremental data computation and view maintenance to efficiently perform the data preparation steps for a new training run.
Indeed, most of the data analysis operators in Section~\ref{sec:data_complexity} lend well to incremental view maintenance techniques~\cite{10.1145/2902251.2902286}.

\subsubsection{Dataset Similarity}
\label{sec:data_sim}
The previous analysis focused on the reuse of the \emph{same} data spans across graphlets. However, what if the data spans are different but their contents have similar distribution? Answering this question can help us identify further opportunities for reuse and optimization, albeit of a different nature.  For instance, if two consecutive graphlets have data inputs with the same distribution, then some aspects of data preparation can be reused even if the data spans are different, e.g., the second graphlet can reuse the vocabularies of the first graphlet over the same categorical features (see also Section~\ref{sec:data_complexity}). One ``extreme'' optimization is to skip training altogether, given that the data distribution is the same!

How can we measure dataset similarity between consecutive graph\-lets? 
As noted above, the input to a graphlet can include multiple data spans. However, we do not have access to the actual data in each graphlet, since this data is not part of our corpus and it might also be siloed for privacy reasons. Instead, the only information we have about each data span is summary statistics for the set of numerical and categorical features in the span (see Section~\ref{sec:data_complexity}). These limitations make it hard to reuse existing data-similarity metrics as is, which either require full knowledge of the data, not just summary statistics, or do not handle comparisons of sets of data spans.

We thus adapt existing metrics to quantify dataset similarity in our specific setup.
We do not claim novelty of this metric; rather, our goal is convey that this
adapted metric provides reasonable
and intuitive results for quantifying
span-content similarity.
A formal investigation on similarity metrics for this type of data is beyond the scope of our work. In what follows, we describe the main intuition behind the proposed metric, leaving the full details to Appendix~\ref{sec:metric}.

The proposed metric has a layered formulation. First, we consider how to compare a pair of data spans using summary statistics on the constituent features. We treat each span as a set of features and use Earth Mover's distance~\cite{rubner2000earth} to compare the two sets, with a feature-to-feature similarity that is based on a hashing scheme on probability distributions~\cite{mao2017s2jsd}. Second, given this span-pair metric as a building block, we introduce a metric to compare two sequences of spans that come from different graphlets. The metric aligns the two sequences by position and computes the sum of span-to-span similarity values normalized by the length of the longer sequence. We use sequences instead of sets to model the sequential visitation of data for certain ML algorithms. If this aspect is not important for a different workload/system, then it is possible to use other metrics such as maximum bipartite matching.

The second row in Table~\ref{tab:similarity_histograms} shows the quartiles of data-similarity values over all pairs of consecutive graphlets in the corpus. Similar to Jaccard similarity (first row), dataset similarity is bimodal at the first and last quantiles. However, the trend is reversed, due to the fact that data spans common in both sets may not be paired up in to ordinal position matching. As explained above, this is by design to account for cases of sequential data visitation in training. The third row in Table~\ref{tab:similarity_histograms} shows the quantiles of data similarity when the latter is averaged over all graphlets in the same pipeline. Compared to the second row, we observe a drop in the higher quantiles, which indicates that pipelines with a large number of graphlets have more dissimilar pairs. Put differently, long-running pipelines belonging to power users have higher data volatility, motivating the need for data validation to safeguard against data errors and drift.

\subsection{Model Retraining and Deployment}
\label{sec:cadences}
In this section, we use the graphlet decomposition to examine the relationship between model (re)training and deployment in the pipelines. As our analysis shows below, the conclusion is that (a) there are many trained models that are not deployed, and (b) very likely they correspond to wasted computation, which in turn introduces an interesting optimization opportunity.

\subsubsection{Training vs. Deployment}
A model {\em push} marks the deployment of a trained model to a downstream service, thus making the model ``visible'' outside of the pipeline. In TFX, a push is performed via the execution of the Pusher operator. Note that a newly generated model may remain unpushed (e.g., due to failure to validate, or downstream throttling), in which case downstream services will keep using the last-pushed model. In this sense, a model push causes a model ``refresh'' for the downstream services.

\begin{figure*}
    \centering
    \begin{subfigure}{0.32\textwidth}
    \centering
    \includegraphics[width=\textwidth]{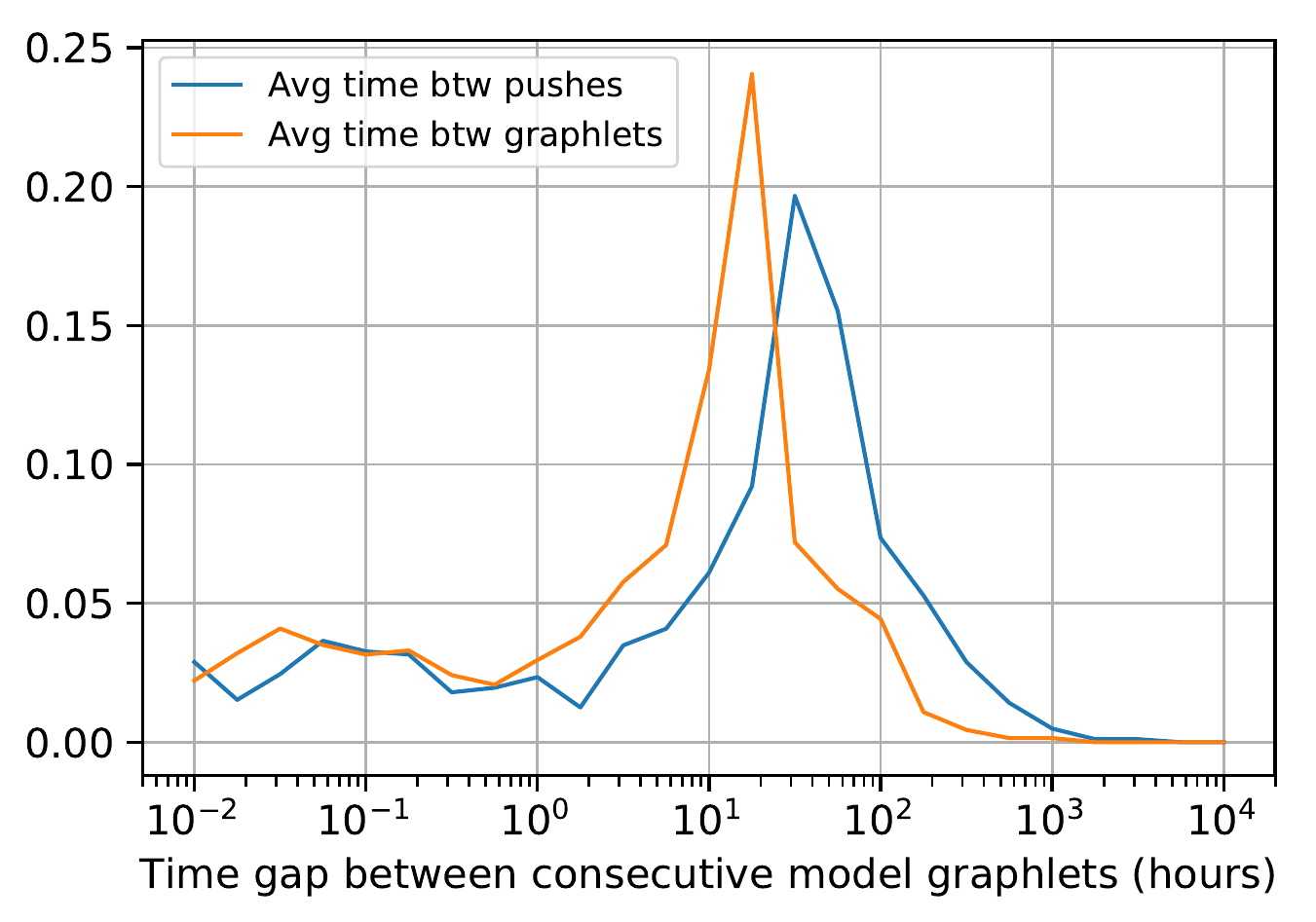}
    \caption{PDF of the average time between consecutive model graphlets.}
    \label{fig:pdf_time_btwn_glets}
    \end{subfigure}\hfill
    \begin{subfigure}{0.32\textwidth}
    \centering
    \includegraphics[width=\textwidth]{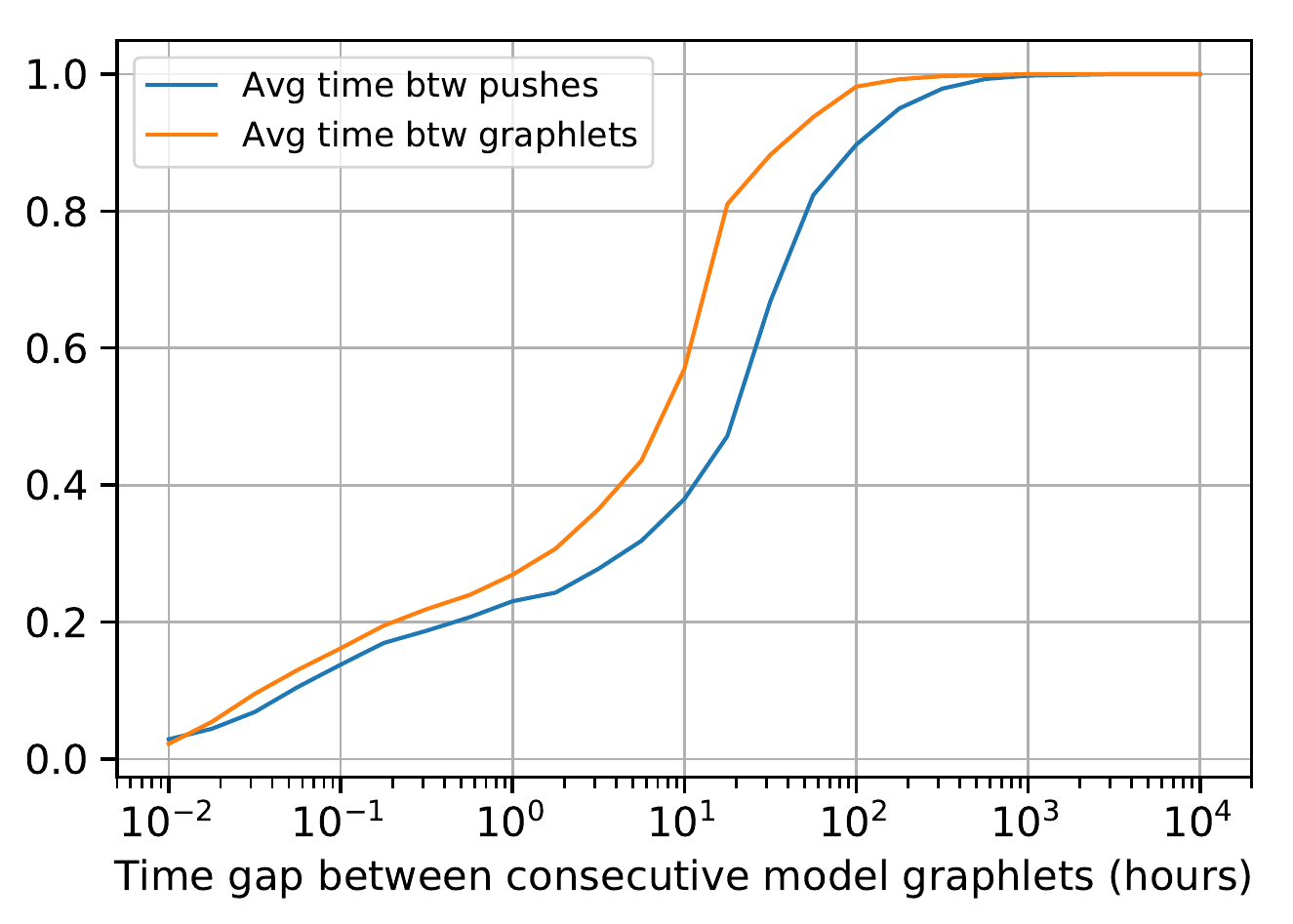}
    \caption{CDF of the average time between consecutive model graphlets.}
    \label{fig:cdf_time_btwn_glets}
    \end{subfigure}\hfill
    \begin{subfigure}{0.32\textwidth}
    \centering
    \includegraphics[width=\textwidth]{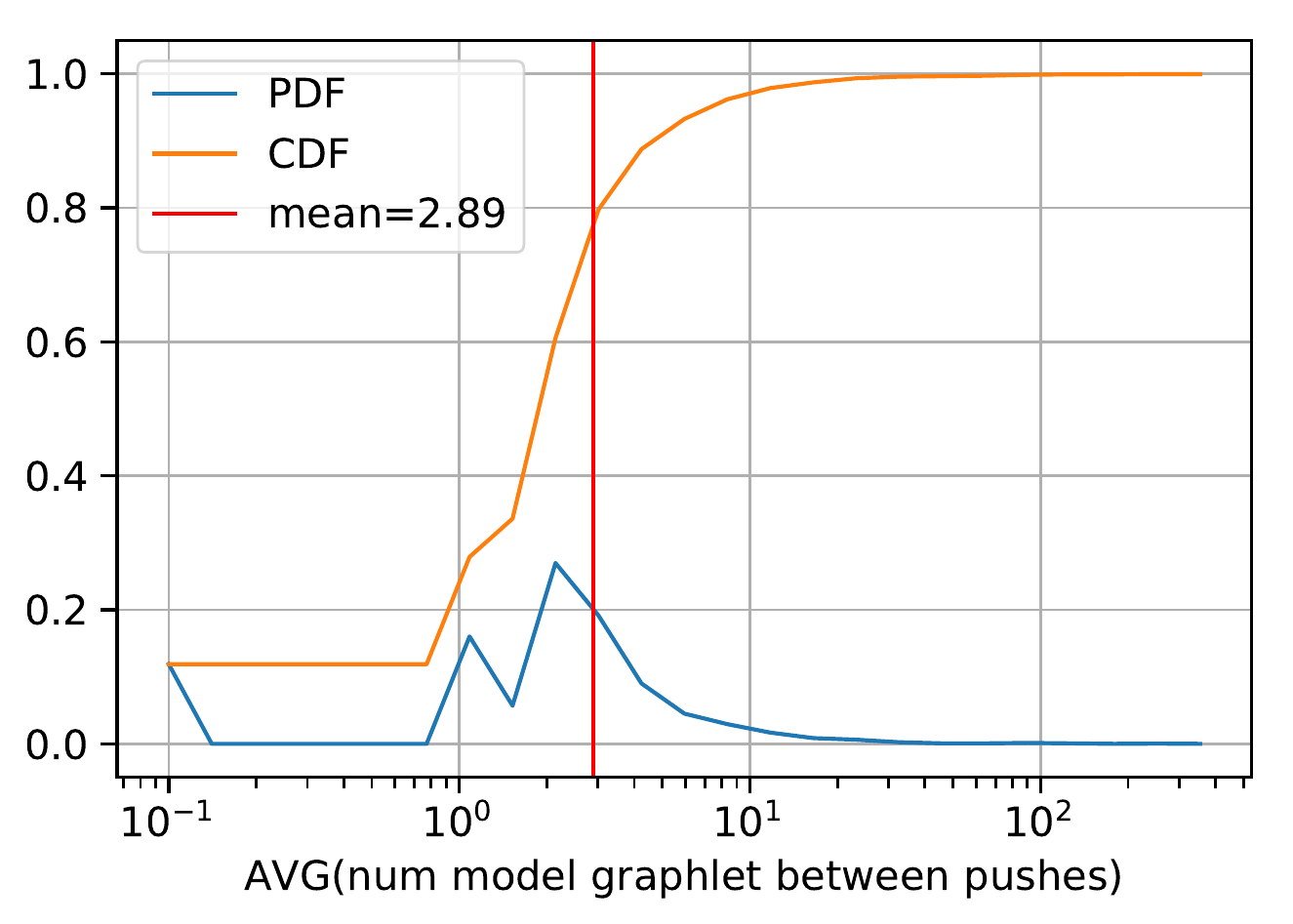}
    \caption{Distribution of number of graphlets between pushes.}
    \label{fig:pdf_cdf_glet_btwn_pushes}
    \end{subfigure}\\
    \begin{subfigure}{0.32\textwidth}
    \centering
    \includegraphics[width=\textwidth]{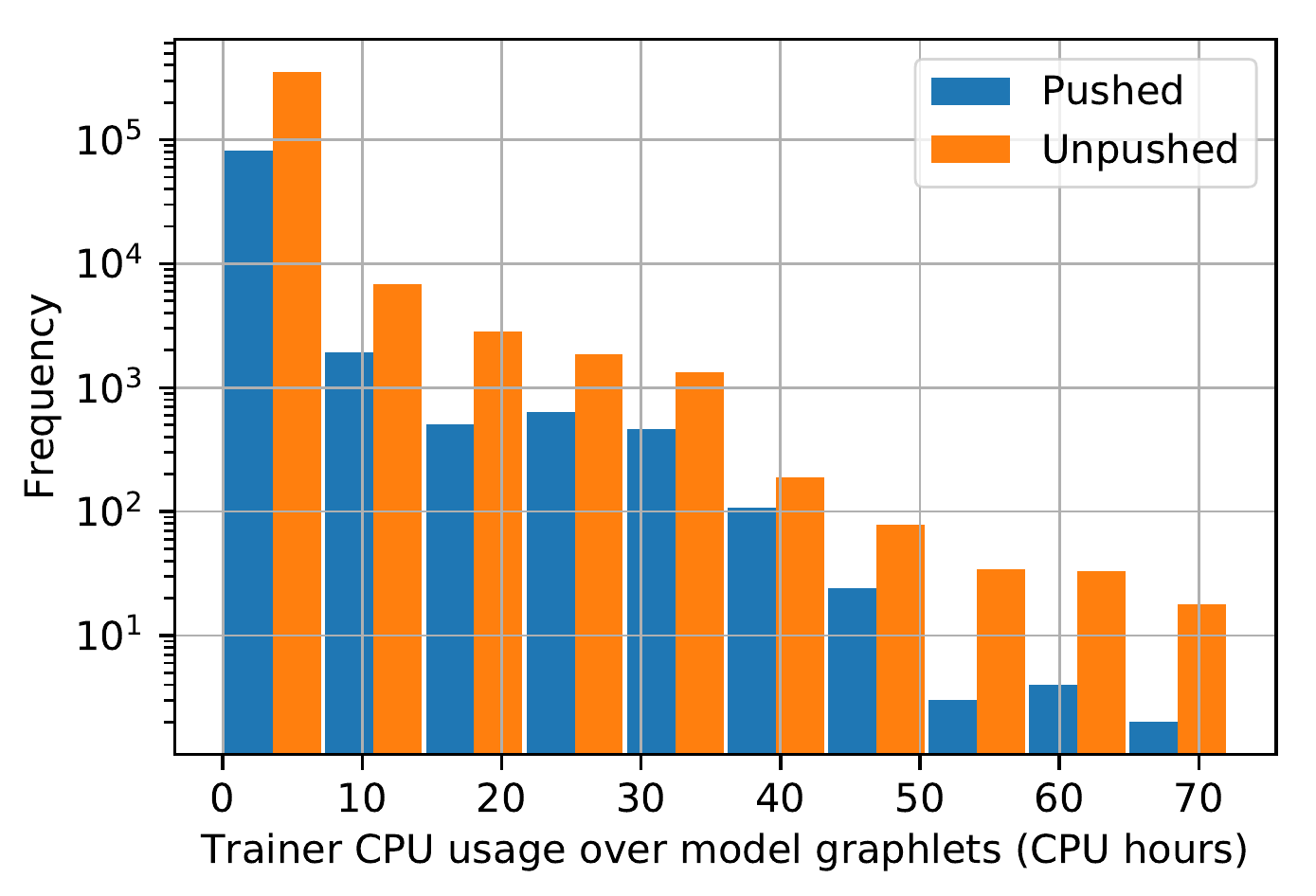}
    \caption{CPU usage for the trainer}
    \label{fig:cpu_pushed_vs_unpushed}
    \end{subfigure}\hfill
    \begin{subfigure}{0.32\textwidth}
    \centering
    \includegraphics[width=\textwidth]{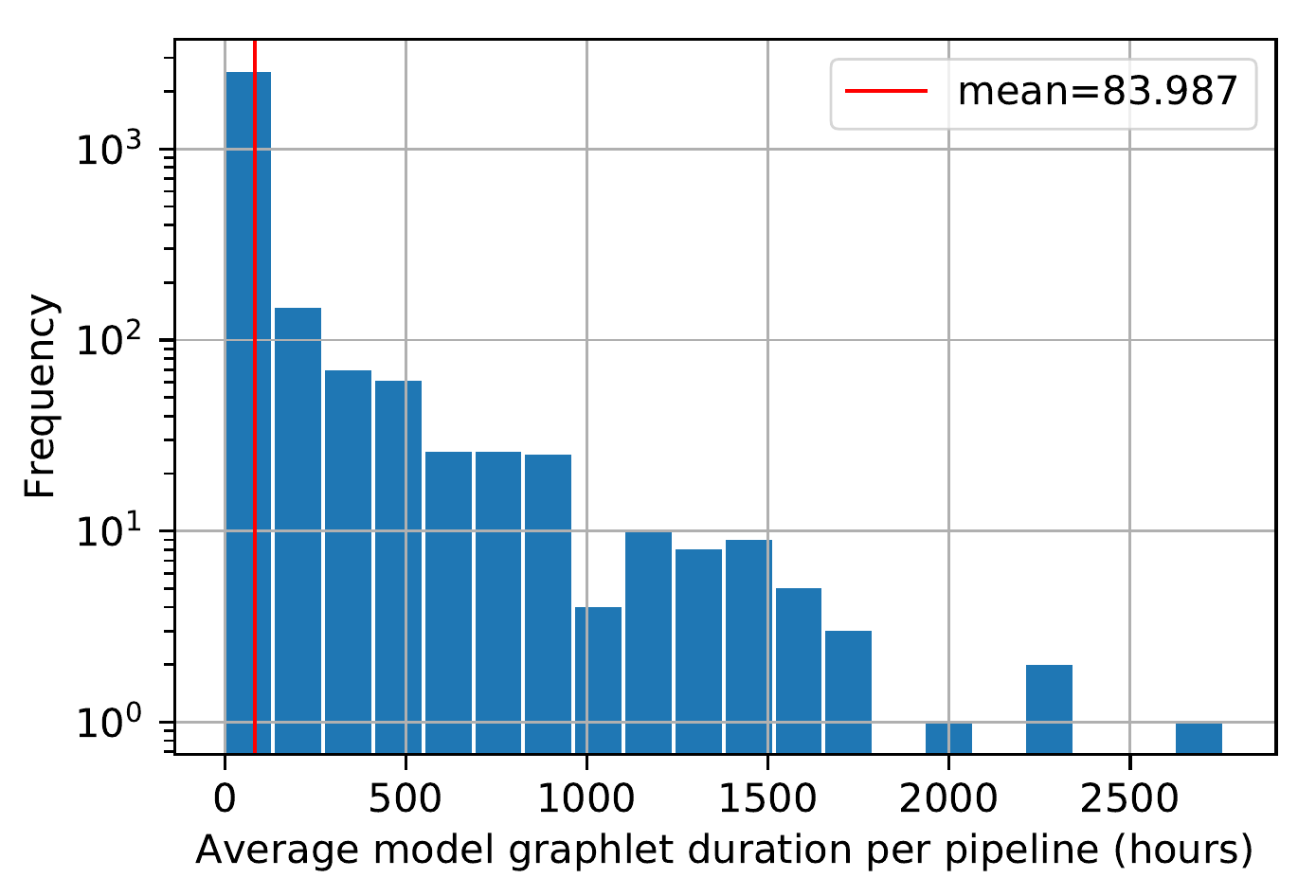}
    \caption{Model graphlets duration in hours.}
    \label{fig:glet_duraction}
    \end{subfigure}\hfill
    \begin{subfigure}{0.32\textwidth}
    \centering
    \includegraphics[width=\textwidth]{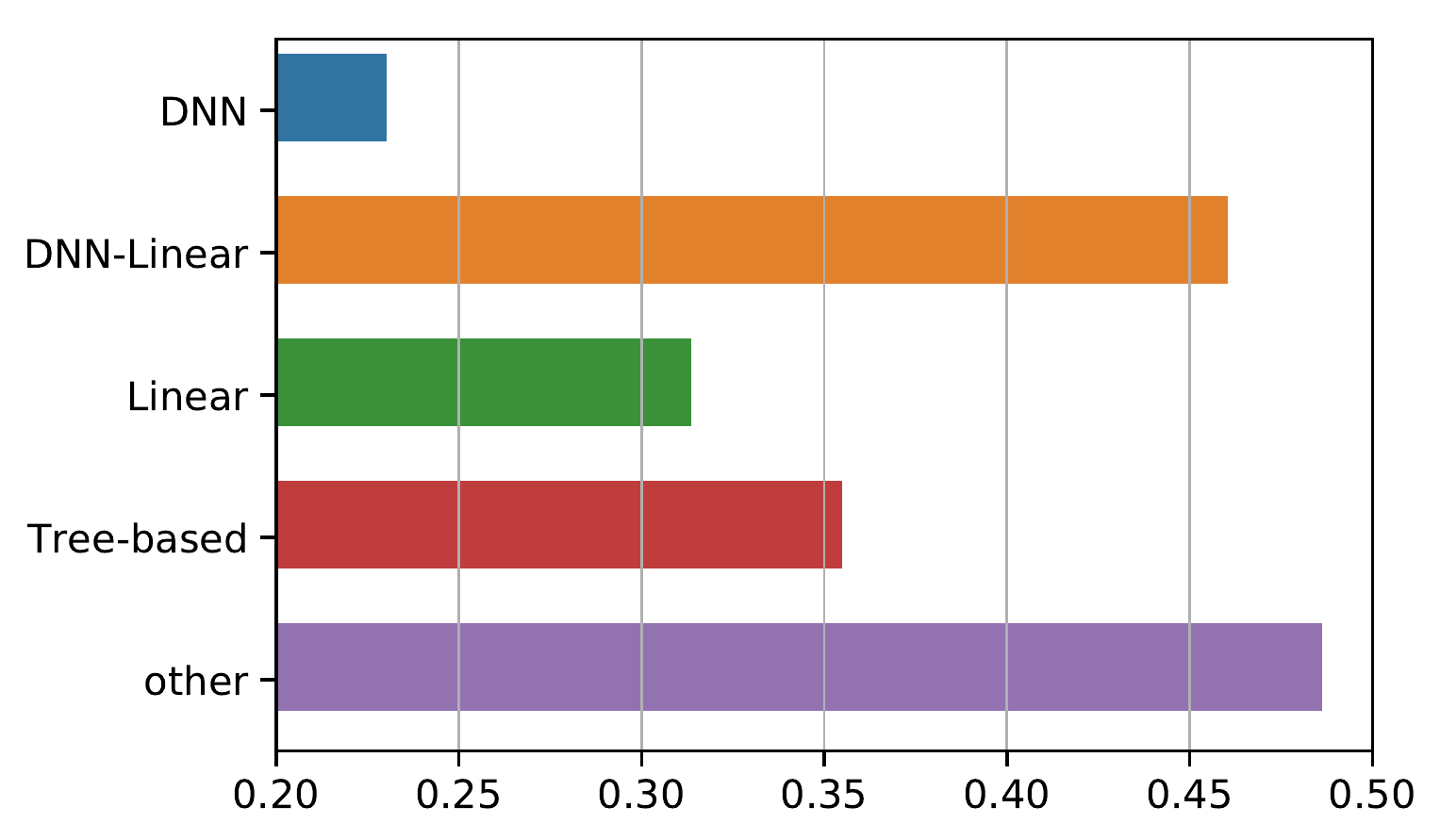}
    \caption{Model type vs. likelihood of pushes.}
    \label{fig:model_push_rate}
    \end{subfigure}
    \caption{Model graphlet analysis.}
    \label{fig:time_btwn_glets}
\end{figure*}

Figure~\ref{fig:pdf_time_btwn_glets} shows the distributions of this time gap (in hours) for the two classes, while Figure~\ref{fig:cdf_time_btwn_glets} shows the cumulative view of the same distributions, both in log scale on the x axis.
Immediately, we see that the two distributions have the same shape, but the mean is upshifted by $\sim$15 hours for the time between pushed graphlets. 
This is corroborated by the CDFs, which show that the time between 80\% of all graphlet pairs is less than the median value for the time between pushed graphlets. 
These trends can have two possible explanations: 1) pushed graphlets tend to run longer than unpushed graphlets, or 2) pushed and unpushed graphlets have similar run times but are interleaved, thus widening the time gap between pushes.
The second explanation has important implications for system optimization, motivating further effort to unearth the cause.

To further investigate the discrepancy between model training cadence and model push cadence, we plot the distribution of the number of graphlets between consecutive model pushes in Figure~\ref{fig:pdf_cdf_glet_btwn_pushes}.
We see that very few pipelines have no intervening unpushed graphlets between consecutive pushes, while most pipelines have between 1 to 10 unpushed graphlets 
between consecutive model pushes, with an average of $\sim$3. 
Furthermore, Figure~\ref{fig:cpu_pushed_vs_unpushed} shows that unpushed graphlets have higher training costs than pushed ones overall.
These two facts together confirm that the second explanation presented above, namely the interleaving of pushed and unpushed graphlets, is indeed the cause for the difference in distribution observed in Figure~\ref{fig:pdf_time_btwn_glets}. 

\subsubsection{Model Freshness vs. Wasted Computation}

The conclusion that pushed graphlets are interleaved with unpushed graphlets is a cause for concern for two reasons:
1) if the application requires model pushes to keep up with the ingestion of new data, then a mismatch between the cadences of model training and model pushing implies an ``unhealthy'' pipeline;
2) if the application is tolerant of unpushed models, model training runs that do not lead to model pushes can be skipped to minimize wasted computation.
In either case, the unpushed graphlets are potentially wasted computation. 

While we do not have the requisite telemetry to precisely characterize when unpushed graphlets are indeed wasteful (e.g., unpushed models might be transitively useful by warmstarting models that do get pushed), 
our analysis results present quantitative evidence that the total amount of wasted computation is high.
First, we observe that approximately 80\% of all graphlets are unpushed. 
If none of the graphlets were used for warmstarting and there were no overlaps between graphlets, 80\% of graphlets would account for $\sim$80\% of overall computation since pushed and unpushed graphlets have similar CPU usage. To account for warmstarting and potential overlaps between graphlets, we can simply 1) remove graphlets belonging to pipelines with warmstarting, which account for 9\% of total graphlets, 
and 2) remove all computation cost for operators that could potentially overlap, which accounts for $\sim60\%$ of overall computation resource consumption, computed by removing the warmstarting pipelines from the results in Figure~\ref{fig:comp_cpu}. Even with these generous assumptions the waste is still at $> 30\%$.

There is clearly an interesting question in whether we can identify the root cause of this inefficiency. However, further analysis shows that it is unlikely to find simple explanations. (We develop a more involved solution for the problem in Section~\ref{sec:waste}.) Specifically, suppose that we attempt to explain the inefficiency through the following hypotheses:
1) rate-limited push: model pushes are configured to be at least a certain time apart, and models are trained faster than the predetermined push rate;
2) model types: certain model types can be more error prone and fail due to non-determinism with all else held equal;
3) data drift: shift in data distribution prevents models from passing validation checks;
4) code change: updating the code for the Trainer introduces system or model bugs.

To check (1), we compare the distribution of the model training time shown in Figure~\ref{fig:glet_duraction} with the distribution of the time gap between graphlets shown in Figure~\ref{fig:time_btwn_glets}. While the average time gap between pushed graphlets is $\sim$40 hours, the mean model training time is 168 hours, making it highly unlikely that models are trained faster than the allowed push rate. This eliminates 1) as the cause of wasted computation.

To check (2), we observe from Figure~\ref{fig:model_push_rate} that the likelihoods of graphlets being pushed for different model types is highly variable.
Moreover, all model types have a likelihood $<0.6$, indicating that no single one is the culprit. However, it is unclear whether the low likelihood for some model types is due to the model type being error prone or other confounding factors. 

\begin{table}[tp!]
    \centering
    \begin{tabular}{c|c|c|c}
           & $\mu_{pushed}$ & $\mu_{unpushed}$ & $\mu$  \\
           \hline
        \textbf{Input data similarity} & 0.109 & 0.099 & 0.101 \\
        \hline
        \textbf{Code match} & 0.838 & 0.846 & 0.845
    \end{tabular}
    \vspace{2mm}
    \caption{Model push vs. data drift and code change.}
    \label{tab:code_data_change}
\end{table}

Finally, to check (3) and (4), we compare the means for input data similarity, as measured by the metric introduced earlier, and code match (where 1 indicates match and 0 indicates no match) with the immediate preceding graphlet for the pushed and unpushed graphlets shown in Table~\ref{tab:code_data_change}. 
Overall, between consecutive graphlets, the input data similarity is 0.101, and the code stays the same 84.5\% of the time.
For both measures, we observe no significant difference between the pushed and unpushed groups. 
These findings suggest that code change and data drift are not major causes of a graphlet not pushing a model.

\newcommand{\yhat}{$\hat{\mathcal{Y}}$\xspace}

\section{Mitigating Wasted Computation}\label{sec:waste}
Our analysis in Section~\ref{sec:fine-grain} revealed that unpushed graphlets represent a significant chunk of the total pipeline computation, even though they are not likely to have an observable impact on subsequent models or downstream services. In turn, this raises a new optimization opportunity: predict whether a graphlet will not lead to a pushed model and then adjust its execution in order to conserve compute cost. For instance, the pipeline scheduler may choose to down-prioritize or stall such graphlets until the pipeline owner intervenes and fixes the underlying issue(s).

In this section we design a decision function for this prediction task. 
Clearly, the decision function has to balance between different types of errors and their effects. A false negative, i.e., skipping a graphlet that would have resulted in a model push, compromises model freshness, while a false positive, i.e., running a model graphlet that does not result in a model push, contributes to wasted computation. We discuss later a method to explore the tradeoff between the two error types and show that we can eliminate nearly half of the wasted computation.

\topic{Data}
To study this problem, we form a dataset by filtering the previous corpus to only include pipelines that do not warmstart model training with previous versions of the model. Unpushed graphlets are useful if they help warmstart subsequent model training (see Section~\ref{sec:cadences}) and so we should not consider them as wasted computation. 
This leaves us with 2827 pipelines containing 420k graphlets in total.
The dataset contains 80\% unpushed graphlets and 20\% pushed graphlets. 
To account for class imbalance, we use balanced accuracy to measure the fitness of decision functions.

\subsection{Problem Statement}
\label{sec:ml_ps}
Let $\mathcal{G} = \{g\}$ be a set of model graphlets and $\mathcal{Y}: \mathcal{G} \rightarrow \{0, 1\}$ be the indicator function that evaluates to 1 for pushed graphlets and 0 otherwise.
The objective of the waste mitigation problem is to find
\begin{equation}
\label{eq:waste}
    \argmin\limits_{\hat{\mathcal{Y}} \in \mathcal{H}} 
    \sum\limits_{g \in \mathcal{G}} 
    L_m \left( \mathcal{Y}(g)\cdot (1 - \hat{\mathcal{Y}}(g)) \right) + 
    L_w \left(\hat{\mathcal{Y}}(g) \cdot (1 - \mathcal{Y}(g)) \right)
\end{equation}
where $\mathcal{H}$ is the space of decision functions that we will explore to find \yhat, an approximation for $\mathcal{Y}$,
$L_m$ is the loss function for model freshness that depends only on false negatives,
and $L_w$ is the loss function for wasted computation that depends only on false positives.
Intuitively, model freshness is only affected by false negatives, where pushed graphlets are incorrectly predicted as \code{unpushed} and therefore prevented from updating the externally visible model; on the other hand, wasted computation can only be incurred by false positives where an unpushed graphlet was erroneously run without resulting in a refreshed model downstream.

The loss functions can be designed to prioritize either cost saving or model freshness.
However, it is difficult, in practice, to determine the tradeoff \textit{apriori} without getting a sense of the complexity of the dataset and the decision function.
To overcome this challenge, one can use a single loss function $L$ for both $L_m$ and $L_w$ to weigh the two types of errors equally and allow \yhat to be a real-valued function that assigns likelihoods to the two labels, namely \code{pushed} and \code{unpushed}. 
Once \yhat is found, the tradeoff between compute cost and model freshness can then be made post-hoc by setting a specific threshold on \yhat to produce a binary decision function. 
Another convenient consequence of setting $L_m = L_w$ is that the problem becomes a standard binary classification problem that can be solved using standard approaches.

\topic{Limitations of Simple Heuristics}
The analyses from Section~\ref{sec:coarse-grain} and~\ref{sec:fine-grain} surface numerous candidate signals that can be used to solve the binary classification problem. 
We experimented with a few simple heuristics for solving Eq.~\ref{eq:waste} derived from our findings, e.g., model type, input overlap, and code match. The best handcrafted heuristic (model type) achieved a balanced accuracy of 0.6. The large search space of heuristics, on top of the low performance of heuristics we handcrafted, motivates the machine learning approach to automatically utilize complex signals for decision making.

\subsection{Machine Learning Based Approach}
\label{sec:ml}
We approximate the decision function \yhat in Equation~\ref{eq:waste}
by training a supervised ML model. 
Each graphlet in the training data 
is labeled as \code{pushed} or \code{unpushed} as described above.
For each graphlet, we create features based on its structure and associated metadata, 
mostly relying on the insights discussed in Section~\ref{sec:coarse-grain} and~\ref{sec:fine-grain}. We also introduce features involving the immediately preceding graphlets in order to capture temporal signals such as data-span similarity. 

\subsubsection{Features}
We partition graphlet features into four categories described below. The first two categories (shape and model information) are features extracted from the graphlet itself. The remaining two categories (input data and code change) are history-based, i.e., they are derived by comparing a graphlet with a window of
graphlets that immediately precede it. A distinct feature is created for each ordinal position in the window, e.g., for a window size of three, \code{code\_change\_1}, \code{code\_change\_2}, and \code{code\_change\_3}, are created to indicate whether the code in graphlet $g$ has changed compared to those that are 1, 2, or 3 graphlets prior to $g$.

\topic{Graphlet shape} 
Shape features include the count of executions corresponding to each operator, as well as the average input and output count for each execution. We partition the operators into \textit{pre-trainer} operators that can execute without the output of the Trainer, the Trainer, and \textit{post-trainer} operators that validate the output of the Trainer for safety and quality.
For example, the graphlet in orange in Figure~\ref{fig:extraction} has 2 ExampleGen with on average 1 output, 1 Trainer with on average 2 input and 1 output, 1 Data Analysis and 1 Model analysis both with average input and output count both being 1, and 1 Pusher with on average 1 input.
Note that to obtain the execution count for a particular operator, we need to run the graphlet up to that operator, incurring computation overhead to obtain these features.
Recall that the additional cost to run the post-trainer operators is $\sim$30\% of the cost to run the pre-trainer operators, as shown in Figure~\ref{fig:comp_cpu}.

\topic{Model information}
We include the model type, e.g. Linear, DNN, etc., as well as the model architecture for graphlets containing DNN models, as one-hot encoded features. Figure~\ref{fig:model_push_rate} 
indicates that model type and model pushes are correlated.

\topic{Input data}
Input data-related features include both 
overlap computed using the input Jaccard similarity (Section~\ref{sec:jaccard-sim}), and
dataset similarity (Section~\ref{sec:data_sim}), 
between a graphlet $g$ and the graphlets preceding $g$ as history-based features.

\topic{Code change}
A binary feature for whether the code versions for the Trainer operator match between a graphlet $g$ and the graphlets preceding $g$, as history-based features. 
While previous results in Section~\ref{sec:ml_ps} have shown that the code change as a standalone feature is low-signal, we include it in the feature set to explore potential interactions with other features.

We also experimented with pipeline-level features inspired by the findings in Section~\ref{sec:coarse-grain}, such as the data transformations,
average time between graphlets and average graphlet duration in a pipeline, 
but found no significant improvement in model performance.

\subsubsection{ML Model Training and Testing}
We split pipelines in the corpus at random into a training and a test set, such that the total number of graphlets belonging to pipelines in the training set is $\sim$80\% of the total number of graphlets in the entire corpus, and the distribution of pushed and unpushed graphlets are roughly the same in the training and test sets.

We experimented with a large variety of models including DNNs and Gradient Boosted Decision Trees, as well as  more interpretable models, such as Logistic Regression and Random Forest, using the Scikit-learn library~\cite{scikit-learn}, and found that Random Forest performed comparably with the more complex models explored by the Auto-ML tool.
We report our results from using the Random Forest model below and discuss lessons learned from the model next.

\subsection{Evaluation}
\label{sec:ml_eval}
We evaluate the models on both model performance and their ability to generate execution policies for reducing wasted computation. 

\subsubsection{Classification performance}
As mentioned above, Random Forest is the most accurate among interpretable models and has comparable performance with much more complex models.  
We present the results for four variants of the Random Forest model: 

\noindent\textbf{RF:Input} has all of the features except the graphlet shape features;

\noindent\textbf{\blue{RF:Input+Pre}} has all of the features in RF:Input plus the graphlet shape features for pre-trainer operators;

\noindent\textbf{\blue{RF:Input+Pre+Trainer}} has all of the features in \blue{RF:Input+Pre} plus graphlet shape features for Trainers;

\noindent\textbf{RF:Validation} has all of the features in \\ \blue{RF:Input+Pre+Trainer} plus shape features for post-trainer operators in the graphlet.

The four variants can be thought of as incrementally 
revealing more features about the shape of the graphlets as more operators are executed. 
They represent points in the pipeline execution that the system can intervene and abort execution to minimize wasted computation.
For example, with RF:Input, the system can decide to abort as soon as the data is ingested; with \blue{RF:Input+Pre}, the system has the option to abort right before model training.
Note that RF:Validation is rather impractical in term of reducing wasted computation, as it requires all operators in the graphlet to be run to obtain the features. Also, the execution of validation operators can be highly correlated with whether the graphlet is pushed. 
We include RF:Validation as a proxy for an 
upper bound on prediction performance given complete (i.e., oracular) information.

\definecolor{Gray}{gray}{0.5}

\begin{table}
    \begin{tabular}{rrll}
    \toprule
            & Model &  Balanced Acc. &  Feature Cost \\
    \midrule
     &   RF:Input &      0.737 &  0.31 \\
        Random                 &      \blue{RF:Input+Pre}  &           0.801 (+9\%) &  0.53 (+71\%) \\
    Forest &      \blue{RF:Input+Pre+Trainer} &             0.818 (+2\%) &  0.77 (+ 45\%) \\
                        & \textcolor{Gray}{RF:Validation} &  \textcolor{Gray}{0.948 (+16\%)} & \textcolor{Gray}{1.00 (+30\%)}\\
                              \hline
    \multirow{4}{*}{Ablation}   & RF:Input &         0.737 &  0.31 \\
                            &  RF:History &         0.738 &  0.77 \\
                        &    RF:Shape &   0.680 &  0.77 \\
                         &    RF:Model-Type &         0.592 &  0.77 \\
    \bottomrule
    \end{tabular}
    \vspace{2mm}
    \caption{Balanced accuracy for all model variants. The feature cost column indicates the compute cost to obtain the necessary features required by the models (rescaled to $[0, 1]$ with RF:Validation = 1).}
    \label{tab:acc_cost}
\end{table}

Table~\ref{tab:acc_cost} shows the model accuracy and the computation cost for the features used in each model, with the cost rescaled to $[0, 1]$ by dividing by the maximum cost attained by RF:Validation.
Adding additional information leads to better model accuracies, 
however, the gain in accuracy does not grow linearly with the compute cost. 
Notably, from \blue{RF:Input+Pre} to \blue{RF:Input+Pre+Trainer}, the difference between the two models corresponds to 
shape features for the Trainer operator, which incurs a 45\% increase in computation cost, for a measly 2\% lift in model accuracy. 
All of the models significantly outperform the heuristics presented in Section~\ref{sec:ml_ps}, suggesting that there are complex interactions between signals that are hard to capture with heuristics looking at one signal at a time.

\begin{figure}[tb!]
    \centering
    \begin{subfigure}[t]{0.35\textwidth}
    \centering
    \includegraphics[width=\textwidth]{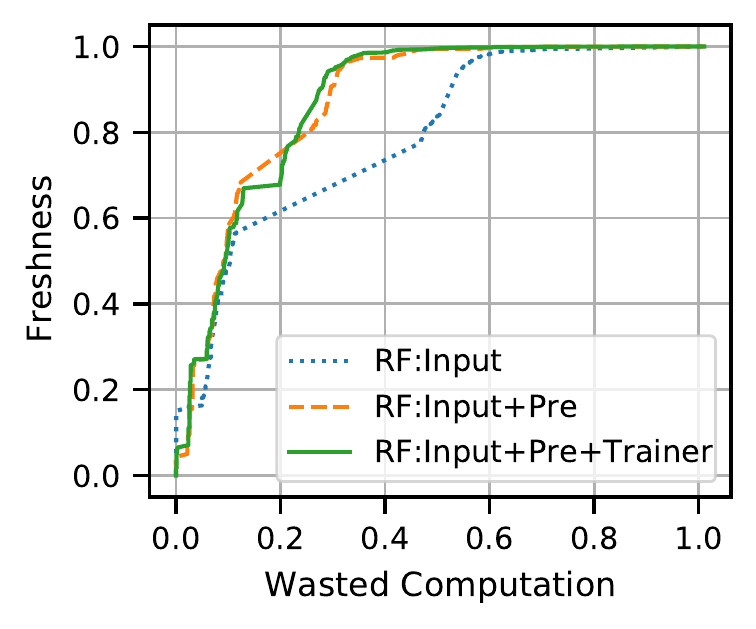}
    \caption{Model freshness vs. wasted computation curves for Random Forest models.}
    \label{fig:waste_short}
    \end{subfigure}\\
    \begin{subfigure}[t]{0.35\textwidth}
    \centering
    \includegraphics[width=\textwidth]{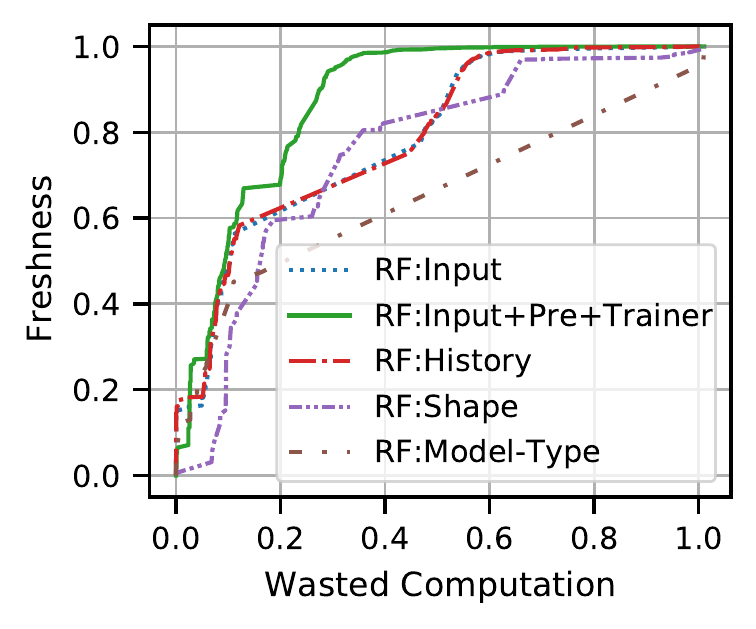}
    \caption{Model freshness vs. wasted computation curves for the ablation study.}
    \label{fig:ablation_sys}
    \end{subfigure}
    \caption{Evaluation Results.}
    \label{fig:evaluation_results}
\end{figure}

\subsubsection{System Performance Improvement}
We now examine how the trained classifier can help balance the tradeoff discussed earlier: skipping unpushed graphlets reduces wasted computation, but skipping pushed graphlets compromises model freshness in downstream applications. 

Figure~\ref{fig:waste_short} depicts this empirical tradeoff between model freshness (y-axis) and wasted computation (x-axis) for the trained classifier. We compute this curve by doing a parameter sweep of the binary-classifier's threshold. Each threshold value results in specific false positive and true positive rates over the evaluation dataset, which we translate to values for wasted computation and model freshness respectively. As an example, consider the rates (0.65, 1) for a specific threshold value, which corresponds to a decision function that classifies correctly all the pushed graphlets and mis-classifies 65\% of the unpushed graphlets (based again on our evaluation dataset). Since all pushed graphlets are identified correctly, model freshness is equal to 100\%. Moreover, we can aggregate the total compute cost of the mis-classified unpushed graphlets, which in this case sums up to 70\% of the total compute cost for these graphlets (equivalently, we recover 30\% of the wasted computation). Hence, we map the decision function to point (0.7, 1) in Figure~\ref{fig:waste_short}.

We highlight two important takeaways from Figure~\ref{fig:waste_short}.
First, we can eliminate 50\% of all wasted computation without sacrificing model freshness at all. 
For a large-scale ML system like TFX, this amounts to a great deal of computation resources saved without affecting end-user experience.
Second, model freshness does not drop substantially if we cut computation even further from 50\% to 60\%. However, following that, model freshness drops dramatically from 0.4 to 0, implying that we cannot hope to do much better than removing 60\% of the wasted compute without significant sacrifice on model freshness.
While RF:Input is cheaper than \blue{RF:Input+Pre} and \blue{RF:Input+Pre+Trainer} as shown in Table~\ref{tab:acc_cost}, it is also a much worse execution policy. It can eliminate at most 30\% of the wasted compute before model freshness suffers from serious decline. On the other hand, \blue{RF:Input+Pre} and \blue{RF:Input+Pre+Trainer} can easily eliminate 50\% of the wasted computation with no impact on freshness.   
Thus, for a frugal user who is tolerant of model staleness, RF:Input might be the preferred model. A user who has strict model freshness requirements might prefer \blue{RF:Input+Pre} to help them minimize waste without sacrificing model quality. The negligible improvement in prediction performance by \blue{RF:Input+Pre+Trainer} compared to \blue{RF:Input+Pre} does not justify the 45\% computation overhead to obtain the features. Thus, \blue{RF:Input+Pre+Trainer}, despite leading in prediction performance, is not as effective from a cost saving perspective.

\subsubsection{Feature Ablation Study}
To better understand the impact of the different groups of features introduced in Section~\ref{sec:ml}, we conduct a feature ablation study, where Random Forest models are trained and evaluated with a subset of the features. 
The balanced accuracies for the ablated models are shown in the last four rows of Table~\ref{tab:acc_cost}.
RF:History contains both the input data features and code change features.
RF:Shape contains only the counts for the operators excluding validators.
Of the four groups of features, RF:Input, while being the cheapest in terms of cost to obtain feature values, achieves the best performing model. Interestingly, adding the code features to the input features, as is done in RF:History, has no effect on the model performance. 
RF:Shape performs significantly worse than RF:Input but better than RF:Model-Type, which achieved the same performance as the simple heuristic involving model type.

Figure~\ref{fig:ablation_sys} shows the freshness vs. wasted computation curves for models in the feature ablation study. 
We use RF:Input+Pre+Trainer as the baseline since it is the best performing Random Forest model.
Of all the feature classes, model features are the least informative by a long shot. 
The fact that RF:Input and RF:History have identical performance serves to validate that code changes are uncorrelated with model pushes. 
The interweaving of the RF:Input and RF:Shape curves suggest that these two groups of features are predictive for different subset of graphlets.
The ablation study shows that no single group of features captures most of the accuracy gains, suggesting that there exist complex interactions between the different groups of features driving the performance of the best models.

\section{Conclusions}
To our knowledge, we presented the first ever large-scale analysis of production ML pipelines, based on a large corpus of pipelines from Google. Our analysis demonstrates the high complexity of these pipelines and the importance of operators other than training, in particular operators related to data preprocessing and analytics. Furthermore, we analyzed the cadence and characteristics of the generated models and characterized their relationship with respect to the input training data. The overall analysis revealed several points where techniques from data management (e.g., incremental view computation, or approximate query answers) can optimize different steps of these pipelines. Moreover, as an example of new optimization problems uncovered by our study, we identified the problem of wasted computation for models that are trained in these pipelines but not deployed to downstream services. To this end, we proposed and evaluated a solution to pro-actively identify such cases and thus avoid wasting resources without compromising the freshness of actually deployed models.

\pagebreak
\bibliographystyle{ACM-Reference-Format}
\bibliography{refs} 

\pagebreak

\begin{appendix}
\section{Graphlet Derivation}
\label{sec:datalog}
Formally, a trace is a directed acyclic graph $G(V, E)$, where $V=\mathcal{A}\cup\mathcal{E}$ includes all artifacts $\mathcal{A}$ and executions $\mathcal{E}$ over time, and $E$ is the union of all input ($\mathcal{A} \times \mathcal{E}$) and output ($\mathcal{E} \times \mathcal{A}$) relations between those operator executions and pipeline artifacts. Given a trainer execution $n\in\mathcal{E}$, a corresponding graphlet $g_n\subseteq G$ is a subgraph of the trace, where its nodes $g_n(V)$ include $n$ and can be derived by the following datalog query:

{\small{
\begin{verbatim}
g(V) :- E(V, X), g(X) 
g(V) :- g(X), E(X, V), NOT sc(V) 
\end{verbatim}
}}
\noindent where \textit{sc} is a predicate that stops to look for more descendant executions related to other trainer executions. In our context, the $sc$ is either Transform or Trainer executions as shown in Figure~\ref{fig:complex_pipeline}.

\section{Input Data Similarity Metric}
\label{sec:metric}
To define the proposed metric, we first consider the case where graphlets contain a single data span as input and then generalize to multiple spans.
A data span $D$ comprises $n$ features $\{f_1, f_2, \ldots, f_n\}$, with each feature being either \textit{categorical} or \textit{numerical}. 
For data privacy concerns, the raw feature statistics are not recorded. 
Instead, for a numerical feature, we have the discrete distribution of the feature values over 10 equi-width bins, with the range rescaled to $[0, 1]$; for a categorical feature, we have the count of the top 10
most frequent terms, the count of the unique terms, and the total number of datapoints, with all terms anonymized. First, we transform the term frequencies for categorical features into a probability distribution by sorting
the normalized term frequencies in the descending order and setting the bin width to be $\frac{1}{N}$, where $N$ is the number of unique terms, to obtain a discrete distribution over $[0, 1]$.
For terms outside of the top 10 most frequent terms, we distribute the remaining mass evenly over the $N -10$ bins. Note that we introduce sorting so that we can capture similarity over the shape of the distribution, independent of the obfuscation of the actual categorical values.

Standardizing feature representation across numerical and categorical features allows us to devise a single unified feature similarity metric for both types of features.
For efficiency, we use a locality sensitive hashing scheme called S2JSD-LSH designed for probability distribution~\cite{mao2017s2jsd} to compute an integer hash value $h(f_i)$ for a feature $f_i$.
The similarity $s(f_1, f_2)$ between two features $f_1, f_2$ with feature names $n_1, n_2$ is then defined as 
\begin{equation}\label{eq:feature_sim}
    s(f_1, f_2) = \alpha \cdot \mathbb{I}(h(f_1) = h(f_2)) + \beta \cdot \mathbb{I}(n_1 = n_2)
\end{equation}
, where $\mathbb{I}$ is the indicator function.
We only compare features of the same type; the similarity between a numerical feature and a categorical feature is always 0.

Given two data spans $D_1 = \{f_i\}$ and $D_2 = \{f'_i\}$, where the $f_i$'s and $f'_i$'s are the features in the respective datasets, the dataset similarity $S(D_1, D_2)$ is defined as the Earth Mover's distance (EMD)~\cite{rubner2000earth} where features are treated as clusters with equal weights within each dataset and the distance between clusters is the feature similarity defined in Eq(\ref{eq:feature_sim}). Since it is difficult to determine a one-to-one mapping between the features given that the feature names are anonymized, we use EMD to capture the uncertainty in the mapping. 
This metric enjoys the property of being symmetric and having a range of $[0, 1]$. Moreover,  $S(\mathbf{\emptyset}, D) = 0$ where $\mathbf{\emptyset}$ is the empty data span, and $S(D, D)=1$.

We extend $S(D_1, D_2)$ to compare two sets of data spans as follows.
Let $\mathcal{D} = (D_1, D_2, \ldots, D_n)$ and $\mathcal{D'} = (D'_1, D'_2, \ldots, D'_m)$ be the sets of input data spans for the two graphlets, where the indices within each set denote order by time of ingestion. 
We define the overall  data similarity as
\begin{equation}
    \label{eq:dataset_sim}
    \mathcal{S}(\mathcal{D}, \mathcal{D'}) = \frac{1}{\max(n, m)} \sum\limits_{i=1}^{\min(n, m)} S(D_i, D'_i)
\end{equation}
Simply put, $\mathcal{S}$ is the sum of pairwise data spans matched by ordinal position, normalized by the maximum number of spans between $\mathcal{D}, \mathcal{D'}$.
We match by ordinal position instead of by identity of the data spans because ordinal positions can better handle rolling windows of data spans that may or may not overlap. Furthermore, certain training algorithms visit data spans sequentially ordered by ingestion timestamp. 
Like $S$, $\mathcal{S}$ is also symmetric and falls into $[0, 1]$. 

\end{appendix}

\end{document}